\newif\ifTwoColumn%
\newif\ifSUBMIT%
\newif\ifCOMMENTS%
\newif\ifFIGs%
\newif\ifFIGoneColumn%
\let\ifSUBMIT\iftrue%
\let\ifCOMMENTS\iffalse%
\let\ifFIGoneColumn\iftrue%

\documentclass[journal=jpcbfk,layout=twocolumn]{achemso}
\setkeys{acs}{articletitle = true}
\mciteErrorOnUnknownfalse
\usepackage{amsmath,graphicx}

\usepackage{array,mathtools,amssymb,booktabs}
\SectionNumbersOn

\ifSUBMIT%
  \ifCOMMENTS%
    \usepackage{color}    
    \usepackage[normalem]{ulem}
    \def\EDITS#1{{\color{red}#1}}
    \def\STRIKE#1{{\color{red}\sout{#1}}}
    
    \def\NSTRIKE#1{{\color{blue}\sout{#1}}}
  \else
    \def\EDITS#1{#1}
    \def\STRIKE#1{}
    
    \def\NSTRIKE#1{}
  \fi
\else
  \usepackage{color}    
  \usepackage[normalem]{ulem}
 \definecolor{mygreen}{RGB}{0,180,0}    
  \def\EDITS#1{{\color{mygreen}#1}}
  \def\STRIKE#1{{\color{red}\sout{#1}}}

  \def\NSTRIKE#1{{\color{blue}\sout{#1}}}
\fi

\usepackage{acronym}

\usepackage{upgreek}

\newlength\figurewide%
\ifFIGoneColumn%
\else
\fi
\title{Invariant Manifolds and Rate Constants in Driven Chemical Reactions}

\author{Matthias Feldmaier}
\author{Philippe Schraft}
\author{Robin Bardakcioglu}
\author{Johannes Reiff}
\author{Melissa Lober}
\author{Martin Tsch\"ope}
\author{Andrej Junginger}
\altaffiliation{Present address: Machine Learning Team at ETAS GmbH (Bosch Group).}
\author{J\"org Main}
\affiliation{%
Institut f\"ur Theoretische Physik 1,
Universit\"at Stuttgart,
70550 Stuttgart,
Germany}

\author{Thomas Bartsch}
\affiliation{%
Centre for Nonlinear Mathematics and Applications,
Department of Mathematical Sciences,
Loughborough University, Loughborough LE11 3TU, United Kingdom
}

\author{Rigoberto Hernandez}
\email{r.hernandez@jhu.edu}
\affiliation{%
Department of Chemistry,
Johns Hopkins University,
Baltimore, \mbox{Maryland 21218, USA}}

\date{\today}

\begin{document}
\newcommand{\EQ}{Eq.}
\newcommand{\EQS}{Eqs.}
\newcommand{\FIG}{Fig.}
\newcommand{\FIGS}{Figs.}
\newcommand{\REF}{Ref.}
\newcommand{\REFS}{Refs.}
\newcommand{\SEC}{Sec.}
\newcommand{\SECS}{Secs.}
\newcommand{\eg}{e.\,g.}
\newcommand{\cf}{cf.}
\newcommand{\ie}{i.\,e.}
\newcommand{\ud}{\mathrm{d}}
\newcommand{\ue}{\mathrm{e}}
\newcommand{\kB}{k_\mathrm{B}}
\newcommand{\VLiCN}{V_\mathrm{LiCN}}
\newcommand{\VCN}{V_\mathrm{C-N}}
\newcommand{\VLi}{V_\mathrm{Li-CN}}
\renewcommand{\vec}[1]{\boldsymbol{#1}}
\newcommand{\qq}{\vec{q}}
\newcommand{\xx}{\vec{x}}
\newcommand{\vv}{\vec{v}}
\newcommand{\transpose}{\mathsf{T}}
\newcommand{\reactantpop}{\mathcal{P}}
\newcommand{\kf}{k_\mathrm{f}}
\newcommand{\etal}{\emph{et al.}}
\newcommand{\LD}{\mathcal{L}}
\newcommand{\LDbo}{\mathcal{L}_\text{bo}}
\newcommand{\LDwo}{\mathcal{L}_\text{wo}}
\newcommand{\LDf}{\LD^\text{(f)}}
\newcommand{\LDb}{\LD^\text{(b)}}
\newcommand{\LDfb}{\LD^\text{(fb)}}
\newcommand{\LDfbw}{\LD^\text{(fbw)}}
\newcommand{\Ws}{\mathcal{W}_\text{s}}
\newcommand{\Wu}{\mathcal{W}_\text{u}}
\newcommand{\Wsu}{\mathcal{W}_\text{s,u}}
\newcommand{\TSt}{\mathcal{T}}
\newcommand{\weightingf}{\chi^\text{(f)}}
\newcommand{\weightingb}{\chi^\text{(b)}}
\newcommand{\weightingfb}{\chi^\text{(f,b)}}
\newcommand{\vtherm}{v_\text{therm}}
\newcommand{\comment}[1]{\textsf{\textcolor{orange}{[#1]}}}
\newcommand{\sno}[1]{_\mathrm{#1}}
\newcommand{\no}[1]{\mathrm{#1}}
\newcommand{\acnew}[1]{\acfi{#1}\acused{#1}}
\newcommand{\VMorse}{V_\text{Morse}}
\newcommand{\VGauss}{V_\text{Gauss}}
\newcommand{\Eg}{E_\text{G}}
\newcommand{\xb}{x_\text{b}}
\newcommand{\subbo}{_\text{bo}}
\newcommand{\subwo}{_\text{wo}}
\newcommand{\xtsr}{\bar{x}}


\begin{abstract}\label{sec:abstract}%
Reaction rates of chemical reactions under
nonequilibrium conditions can be determined through
the construction
of the normally hyperbolic invariant manifold (NHIM) [and moving
dividing surface (DS)] associated with the transition state trajectory.

Here, we extend our recent methods
by constructing
points on the NHIM accurately even for multidimensional
cases.
We also advance the implementation of machine learning
approaches to construct smooth versions of the NHIM from a known
high-accuracy set of its points.
That is, we expand on our earlier use of
neural nets, and introduce the use of Gaussian process regression
for the determination of the NHIM.
Finally, we compare and contrast all of these methods for a
challenging two-dimensional model barrier case so as to illustrate their
accuracy and general applicability.
\end{abstract}
\maketitle

\acrodef{DS}{dividing surface}
\acrodef{TST}{transition state theory}
\acrodef{TS}{transition state}
\acrodef{LD}{Lagrangian descriptor}
\acrodef{LDDS}{Lagrangian descriptor dividing surface}
\acrodef{PODS}{periodic orbit dividing surface}
\acrodef{NHIM}{normally hyperbolic invariant manifold}
\acrodef{TD}{time descriptor}
\acrodef{GP}{Gaussian process}
\acrodef{GPR}{Gaussian process regression}
\acrodef{NN}{neural network}


\section{Introduction}

The framework of \ac{TST} provides a powerful basis for the qualitative and
quantitative description of chemical reactions.
When their dynamics can be described by a Born-Oppenheimer potential
driven by a classical equation of motion,
then \ac{TST} reduces the dynamical calculation to a geometric
one involving the identification of a barrier region
that separates reactants and products.
The
position of the barrier is typically marked by a saddle point of rank~1 that has
exactly one unstable direction which coincides with the \emph{reaction
coordinate} at that point.
The remaining degrees of freedom are locally stable and are associated with
the \emph{bath coordinates.}
In this paper, we aim to
address activated processes with arbitrary dimensionality
but restricted to reactions that can be characterized
by a one-dimensional reaction coordinate.
We do not solve this class of problems in fullest generality, but do
make progress in treating systems with bath coordinates of dimensions
higher than one, addressing both theoretical and numerical challenges
to the computation of the dividing surface.

\ac{TST} rests on the identification of a \ac{DS} in the barrier region
which separates reactants from
products.\cite{eyring35,wigner37,pech81,truh96,peters14a}
It is exact, if the \ac{DS} is crossed once and only once by each reactive
trajectory.
Recrossings of the \ac{DS} lead to an overestimation of the rate.
Advances in the determination of a recrossing-free
\ac{DS} will therefore impact a broad range of problems
beyond those of chemical reactions in which the overall process
can be described as an activated process that is primarily
characterized by a single, albeit curved, reaction coordinate in the
presence of a bath.
Examples abound in atomic
physics~\cite{Jaffe00}, solid state physics~\cite{Jacucci1984}, cluster
formation~\cite{Komatsuzaki99, Komatsuzaki02}, diffusion dynamics~\cite{toller,
voter02b}, cosmology~\cite{Oliveira02}, celestial mechanics~\cite{Jaffe02,
Waalkens2005b}, and Bose-Einstein condensates,\cite{Huepe1999, Huepe2003,
Junginger2012a, Junginger2012b, Junginger2013b}
to name a few.

Of course, the primary aim for the current work is the
construction of accurate rates in chemical reactions that
take place in solution, perhaps under nonequilibrium conditions,
and for which we ultimately want to achieve some level of control.
The basic (or naive)
theory uses a planar dividing surface to determine the flux,
but it is approximate because it is invariably recrossed by trajectories
that go from reactants to products.\cite{eyring35,wigner37,pech81,truh96}
It is also an upper bound, and leads to a
variational transition state theory\cite{truh84,tal05}
which optimizes the dividing surface by
way of minimizing recrossings.
The development of semiclassical transition state
theory\cite{mill77,mill90b,hern93b,mill98b,Stanton2011}
taking advantage of Keck's phase space representation of \ac{TST}\cite{keck67}
hinged on the development of good action-angle variables associated with
the \ac{DS}.
The use of perturbation theory to construct these objects was found in
parallel to work in dynamical systems theory
addressing activated escape.\cite{bristol1,bristol2}
The theory was then generalized for chemical reactions under
time-dependent conditions ---arising from driving, noise, or
both.\cite{dawn05a,hern08d}
Applications to chemical reactions have included
H $+$ H$_2$,\cite{hern94,Allahem12}
LiCN,\cite{hern16c}
and
ketene isomerization.\cite{hern16d}
While these are interesting cases, they do not represent the bulk of chemical
reactions which are generally higher dimensional and which take place in more
complex environments.
This work is thus focused on advancing the theoretical and computational machinery
that have the potential of addressing such challenging systems.

In a multidimensional autonomous Hamiltonian system, a recrossing-free
\ac{DS} is attached to the \ac{NHIM}.
The latter contains all trajectories that are trapped
in the saddle region both forward and backward in time.
It can be constructed
using a normal form
expansion.\cite{pollak78,pech79a,hern93b,hern94,Uzer02,Jaffe02,
Teramoto11,komatsuzaki06a,Waalkens04b,Waalkens13}
For a multidimensional
time-dependent system that is driven by an external field or subject to
thermal noise, the question arises whether the NHIM can be generalized
in such a way that it can still be used to construct
a recrossing-free \ac{DS}.
Recent successful constructions\cite{hern15e,hern17h,hern18c}
of such a \ac{DS}
suggests that it is indeed possible to do so.
The purpose of this paper is to clarify the geometry that
underlies this construction,
and to present a readily applicable procedure to
calculate rate constants of time-dependent, driven and multidimensional
systems
with a rank-1 saddle and an arbitrary number of bath modes.
As this work is focused on advancing the methods so as to ultimately treat
higher dimensional chemical reactions,
numerical results are restricted to model systems ---inspired by
chemical systems--- useful for
verifying the theory.

The computational task is challenging because of the dimensionality of the
objects: In an $n$-dimensional system with one reaction coordinate and $n-1$
bath coordinates, the phase space has dimension $2n$. The \ac{DS} is a
time-dependent and non-trivially curved hypersurface of dimension $2n-1$,
and the \ac{NHIM} is a surface of dimension $2n-2$.
Previous work has shown that attaching
a \ac{DS} to the \ac{NHIM}, whose dimension is lower, can yield a
recrossing-free \ac{DS} if the reaction coordinate is
unbound.\cite{hern17h,hern18c}

An individual point on the \ac{NHIM} can be computed with the algorithm
discussed in Ref.~\citenum{hern18g}.
However, to construct a \ac{DS} that
covers a sufficiently large region in phase space and time, a vast number of
points is required.
Even with an efficient algorithm, computing the required
number of points on the \ac{NHIM} is a daunting task.
Consequently, a continuous interpolation between
the finite collection of
high-accuracy ---and therefore computationally expensive---
points on the \ac{DS} in space and time is needed
to determine the reactivity of a given trajectory.
We present two tools developed in the context of machine
learning to approximate any function between isolated points in
arbitrary high dimensions.
It is notable that Pozun \etal\cite{henkelman12}
earlier implemented a support-vector machine to identify
a \ac{DS} between reactants and products on a time-independent
potential energy surface.
Here, we focus on time-dependent potentials, often coupled to a bath, in which the
associated \ac{NHIM} is consequently time-dependent.
First, a \ac{NN}\cite{FUNAHASHI1989183} can be trained on an
arbitrary set of points located on the \ac{NHIM}.
Once trained, these networks
provide fast access to any point needed for calculating rates in
time-dependent driven systems.\cite{hern18c}
We also present
an alternative machine learning algorithm based
on \ac{GPR}.\cite{murphy2012machine,rasmussen2006gaussian}
We compare these two algorithms
and discuss the relative advantages and disadvantages in their application to
\ac{TST}.

Perturbative\cite{hern06d,Kawai2009a,hern14b,hern14f,hern15a}
and nonperturbative\cite{hern15e,hern16d,hern16a,hern16h,hern16i}
constructions
of a recrossing-free \ac{DS} in time-dependent systems
rely on the concept of the \ac{TS} trajectory:\cite{dawn05a,dawn05b}
a unique trajectory bound to the vicinity of the saddle for all time.
In systems with more than one degree of freedom,
such a \ac{TS} trajectory
is unique only if the system is dissipative
with the oscillations of the bath modes damped asymptotically.
If the reactive system is Hamiltonian
---that is, when in the absence of damping,--- all trajectories on the
\ac{NHIM} are trapped near the barrier for all time
and can therefore serve as \ac{TS}~trajectories.
The consequences of the non-uniqueness of the
\ac{TS} trajectory were avoided in Ref.~\citenum{Kawai07} by an arbitrary
choice of one of the trapped trajectories as ``the'' TS~trajectory.
The \ac{DS} constructed there and in Refs.~\citenum{hern17h,hern18c}
also avoids such arbitrary
choices because it considers the NHIM itself, rather than a particular
trajectory on it.
For multidimensional
time-periodically driven Hamiltonian systems,
we suggest
identification of the \ac{TS} trajectory to
be restricted to the one trajectory,
among all the bound trajectories on the \ac{NHIM},
that is periodic.
This reduces to the usual result in one dimension, and is no longer
arbitrary in higher dimensions as it is now selected by a
characteristic property of the system.

The outline of the paper is as follows:
In Sec.~\ref{sec:TST_theory}, we start
with a brief overview describing how rates in time-dependent systems
with a rank-1 saddle
are calculated using the \ac{NHIM} and the associated \ac{DS}.
This suggests a restricted
definition of the \ac{TS} trajectory stated above.
In Sec.~\ref{sec:points_on_nhim}, we
present a fast and robust algorithm for
calculating single points on the \ac{NHIM}.
Sec.~\ref{sec:machine} summarizes two
interpolation algorithms based on machine learning,
and provides an analysis of the respective
advantages and disadvantages.
In Sec.~\ref{sec:application}, we apply both
methods to calculate rate constants for a two-dimensional system.

\section{TST in time-dependent systems}\label{sec:TST_theory}

A possible first step in obtaining the time-dependent representation of the \ac{DS}
is the determination of individual points on this high-dimensional object.
As outlined in
Ref.~\citenum{hern17h}, in time-dependent systems
it suffices to calculate the \emph{normally hyperbolic invariant manifold}~(NHIM)
because the \ac{DS} in time-dependent systems is attached
to it in a prescribed way.
In this section, we describe the former as a
fundamental geometric object and
demonstrate how it can be used to obtain a \ac{DS}.
We start by considering a one-dimensional system in which
there are no bath coordinates.
We then present the \emph{binary contraction method}
for finding the \ac{NHIM} based on the
geometric properties of phase space in the vicinity of a rank-1 saddle.
Furthermore, we describe how the
insights gained in the one-dimensional setting can be generalized to higher
dimension.
Finally, we will discuss how a unique \ac{TS} trajectory can be
defined in a multidimensional system.

\subsection{The role of the NHIM in TST}
\label{subsec:role_of_the_nhim}

\begin{figure}[t]
  \centering
  \includegraphics[width=0.7\figurewide]{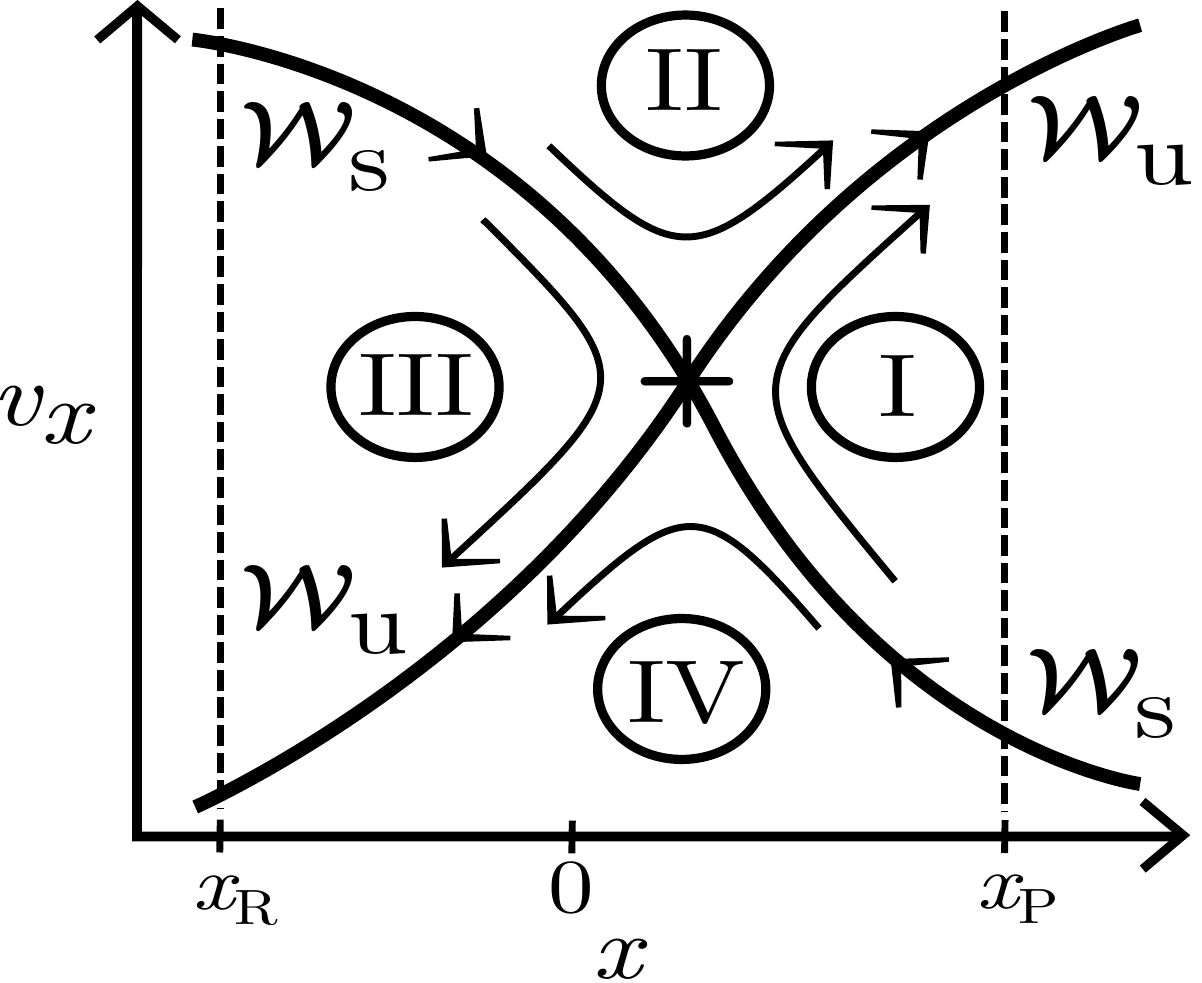}
  \caption{Phase space $(x, v_x)$ close to the boundary point (+) of the stable
    $\mathcal{W}\sno{s}$ and unstable $\mathcal{W}\sno{u}$ manifold. The areas
    of non-reactive trajectories (I) and (III) are separated from the areas with
    reactive trajectories (II) and (IV) by the manifolds. A reaction can either
    take place from the reactant side $x\sno{R}$ to the product side $x\sno{P}$
    or vice versa. The arrows illustrate the heading of the trajectories close
    to the boundary point.
  }
  \label{fig:reactive_regions}
\end{figure}

In a one-dimensional system, a rank-1 saddle is simply a maximum of the
potential.
The relevant phase space structures in its neighborhood are shown in
Fig.~\ref{fig:reactive_regions}: Trajectories that approach the energy barrier
with sufficiently high energy (from either side) will cross the barrier,
and others
will not.
Its phase space can be separated into four different regions
as characterized by where particles came from and where they go to.
The dividing lines between these regions are the stable $\mathcal{W}\sno{s}$ and
unstable $\mathcal{W}\sno{u}$ manifolds. They contain trajectories that are
trapped near the saddle in forward ($\mathcal{W}\sno{s}$) or backward
($\mathcal{W}\sno{u}$) time, respectively.

\begin{table}[t]
  \centering
  \caption{Nomenclature of the different reactive (II, IV) and non-reactive (I,
    III) areas in phase space separated by the two mani\-folds. The particle
    leaves the neighborhood of the saddle at the exit point
    $x\sno{exit}^{\mathrm{b,f}}$ when propagated in backward~(b) and forward~(f)
    time, respectively.
  }
  \label{tab:regions}
  \begin{tabular}{crr}
    \toprule
    Area & $x\sno{exit}^\mathrm{b}$ & $x\sno{exit}^\mathrm{f}$\\
    \midrule
    I & $x\sno{P}$ & $x\sno{P}$\\
    II & $x\sno{R}$ & $x\sno{P}$\\
    III & $x\sno{R}$ & $x\sno{R}$\\
    IV & $x\sno{P}$ & $x\sno{R}$\\
    \bottomrule
  \end{tabular}\label{areas}
\end{table}

A trajectory starting at a given phase space point
is propagated in forward (f) and backward (b) time
until it leaves the saddle region
given by $x\sno{R} \le x \le x\sno{P}$ with a suitably chosen
$x\sno{R}$ at the reactant side and $x\sno{P}$ at the product side of the reaction
coordinate.
The initial point is then assigned to a region, depending on whether it
leaves the saddle region at $x\sno{R}$ or $x\sno{P}$
in forward or backward time, as explained in
Table~\ref{areas}.
Note that numerically, the trapping
cannot, in general, be observed because the dynamics near the saddle is
unstable. Small numerical errors cause an otherwise trapped trajectory to leave
the saddle region in either time direction.

The closures of stable and unstable manifolds intersect in a point (see black
cross in Fig.~\ref{fig:reactive_regions}).
When the potential is also time-independent,
this intersection is a fixed point.
In time-dependent (or driven) cases, it becomes a trajectory that is
trapped in both forward and backward time.
This trajectory
is referred to as the \ac{TS}~trajectory~\cite{dawn05a,dawn05b,
hern06d,Kawai2009a,hern14b,hern14f,hern15a,hern16a,hern16h, hern16i}.
In one-dimensional systems, it is unique.
In multidimensional systems, the initial conditions of trajectories
trapped in both forward and backward time form the time-dependent
multidimensional \ac{NHIM}\cite{Lichtenberg82, Ott2002a,wiggins2013normally}.
The latter reduces to a single trajectory in one-dimensional systems,
and so the unique \ac{TS} trajectory in one dimension
can also be called the \ac{NHIM}.

For a time-independent, one-dimensional system, the \ac{NHIM} is
located exactly at the barrier top.
In a time-dependent system, all structures shown in
Fig.~\ref{fig:reactive_regions} depend on the initial time at which the
trajectories are started, i.e., the stable and unstable manifolds, as well as
the \ac{NHIM}, are themselves time-dependent.
The \ac{NHIM} can be expected to
reside close to the saddle, but it will in general not coincide with it.
If the external driving is periodic in time, the \ac{NHIM} is a periodically moving
manifold with the same period as the external driving force
(see blue, dashed circle in Fig.~\ref{fig:crosses}(d), calculated for the example of
Eq.~\eqref{1d_model}).

So far, we have assumed open reactant and product basins.
In the more typical case of closed basins ---as in e.\,g.~the model potential of
Ref.~\citenum{hern17e}, or potentials of real systems, for instance
LiCN~\cite{hern14j, hern16c, hern12e} or ketene~\cite{hern13c, hern13e, hern14e,
hern16d}---
the requirement for trajectories to be \emph{recrossing-free}
often needs to be restricted to mean \emph{locally
recrossing-free}, i.e., a particle that enters the neighborhood of the saddle
will cross the \ac{DS} no more than once before it leaves the saddle region.
It may, however, reenter this region and recross the \ac{DS} at a later time.
For further information about these \emph{global recrossings} and the
degree to which \ac{TST} can remain exact despite them, see
Ref.~\citenum{hern17e}.

\subsection{Recrossing-free \ac{DS} in time-dependent systems}
\label{subsec:RecrossFreeDS}\label{sec:definition}

For a system with $n$ degrees of freedom, the separation of the $2n$-dimensional
phase space into reactant and product regions requires a surface of dimension
$2n-1$.
To distinguish reactant and product regions for any rank-1 saddle in an
$n$-dimensional energy surface we use an appropriate coordinate system where $x$
approximates the reaction coordinate, i.e., the unstable direction of the
saddle, and $\vec{y}$ the $n-1$ remaining (bath) coordinates, given by the
stable directions of the saddle. The corresponding velocities are denoted by
$v_x$ and $\vec{v_y}$, respectively. We will construct a dividing surface at an
instantaneous reaction coordinate $x^{\mathrm{DS}}(\vec{y}^{\mathrm{p}},
\vec{v_y}^{\mathrm{p}}, t)$ that depends on the bath coordinates, their
velocities, and on time. A phase space point $(x,\vec y, v_x, \vec{v_y})$ is
classified as lying on the reactant or product side of the barrier at time~$t$
depending on whether it fulfills
\begin{align}
\begin{aligned}
    x &< x^{\mathrm{DS}}(\vec{y}^{\mathrm{p}}, \vec{v_y}^{\mathrm{p}}, t)\rightarrow~\mathrm{reactant, or}\\
x &> x^{\mathrm{DS}}(\vec{y}^{\mathrm{p}}, \vec{v_y}^{\mathrm{p}}, t)\rightarrow~\mathrm{product}\,.
\end{aligned}
\label{react_prod}
\end{align}
While propagating an ensemble of particles, each particle can be classified
as reactant or product at any time according to Eq.~\eqref{react_prod}. When it
crosses the moving \ac{DS}, it reacts and contributes to the reaction of
the time-dependent system.

To satisfy the recrossing-free requirement,
this high-dimensional DS
is generally non-trivially curved in a manner that depends on the details
of the dynamics.
The representation~\eqref{react_prod} of the \ac{DS} assumes that no curvature
in the $v_x$ direction is required.
This assumption is inspired by the phase space plot in Fig.~\ref{fig:crosses},
where such choice of a \ac{DS} appears as a vertical line going up from the \ac{NHIM}.
Reactive trajectories cross the \ac{DS} transversely and exactly once,
whereas non-reactive trajectories do not cross it.
This statement is trivially true in a harmonic system.
It describes a qualitative property that is robust under perturbations
of either the system or the dividing surface.
Many other surfaces that are sufficiently close to the vertical line
will also be recrossing-free;
the exact choice of \ac{DS} is therefore not critical.
We take advantage of this non-uniqueness by postulating that
the \ac{DS} has the form~\eqref{react_prod} and that
its location $x^{\no{DS}} = x^{\no{NHIM}}$ coincides with that of the \ac{NHIM}.
These assumptions single out a unique \ac{DS} and also provide a means to compute it
because the NHIM is amenable to numerical computation,
as discussed below in Sec.~\ref{sec:points_on_nhim}.

With this simplification,
our choice of a recrossing-free DS, independent on the reactive velocity $v_x$,
can be justified
via a confirmation of the
absence of unwanted recrossings in dynamical simulations,
as shown in Sec.~\ref{sec:application}
for a particular two-dimensional model system.
However, in more general cases when
particles cross the DS far away from the NHIM,
this simplification might lead to unwanted recrossings (see,
e.~g.~Refs.~\citenum{Komatsuzaki99a,komatsuzaki06b,Li09,komatsuzaki06a}),
as the assumption embodied in Eq.~\eqref{react_prod} has its limits.
There is no general proof that a recrossing-free \ac{DS} of the form \eqref{react_prod}
should exist,
and exceptions are known.\cite{Komatsuzaki99a}
If the invariant manifolds are strongly deformed, e.\,g.~such that they cross the chosen \ac{DS},
then the fate of a trajectory starting on the \ac{DS} $x=x^{\mathrm{DS}}(\vec{y}, \vec{v_y}, t)$
will depend on $v_x$,
and the representation~\eqref{react_prod} becomes impossible.
However, this is expected only for strongly non-linear systems
or very high velocities,
so that the representation~\eqref{react_prod} is adequate for many systems
of practical relevance.
For a given system, one can always numerically confirm
whether a \ac{DS} is truly recrossing-free.
This may be done by propagating a large number of trajectories with
initial conditions in the vicinity of the moving saddle,
and subsequent monitoring of the
crossings across the constructed \ac{DS}, as has been done successfully for the
model system discussed in Refs.~\citenum{hern17h, hern18c}.

\begin{figure}[t]
  \centering
  \includegraphics[width=0.4\figurewide]{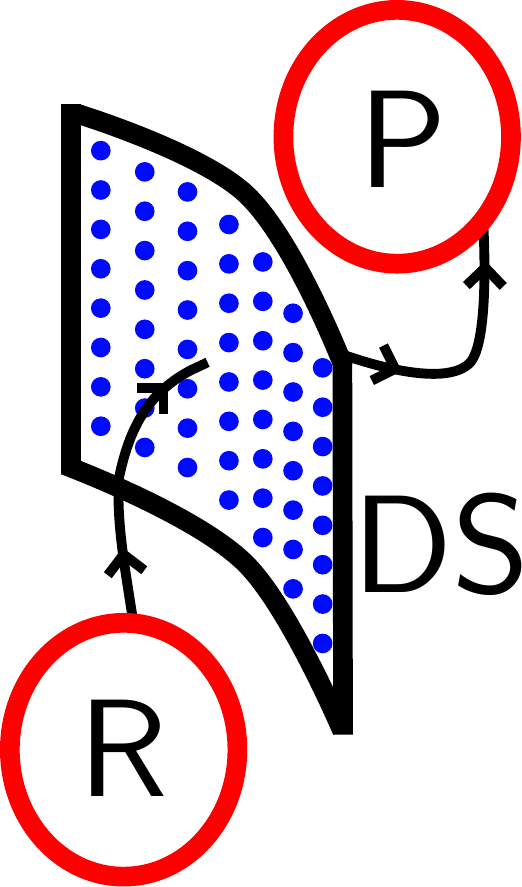}
 \caption{Simplified sketch of the phase space of a chemical reaction. The blue dots
    represent pre-calculated points on the \ac{DS}. A reacting particle
    crosses the \ac{DS},
    a curved, co-dimension one manifold
    separating reactants (R) and products (P) in phase space.
    Here, we illustrate the problem, that such a reacting trajectory will in
    general not cross
    the \ac{DS} at the pre-calculated
    points but somewhere in between. Note, that
    these manifolds
    are generally
    high-dimensional objects embedded in phase space and
    this figure has to be understood as highly schematic.
  }
  \label{fig:sketch_reaction}
\end{figure}

If the underlying potential has open reactant and
product basins, a particle that
crosses a recrossing-free \ac{DS}
will never return because it
escapes the saddle region forever.
Assuming that our choice of a \ac{DS} independent on the
reactive velocity $v_x$ is sufficiently accurate
The propagation of a reacting particle can be stopped
when a particle crosses a space determined by
$x$ = $x^{\mathrm{DS}}$ while calculating rates in open systems.
In systems with closed reactant or product basins,
particles can be reflected globally at the boundaries
of the basins and may therefore
cross the space
determined by $x=x^{\mathrm{DS}}$
infinitely many
times.\cite{hern17e,hern14j, hern16c, hern12e, hern13c, hern13e, hern14e,
hern16d}. In this case, the propagation cannot be stopped and particles can
react from one side of the saddle to the other multiple times. Still, the
time-dependent classification in reactant and product is always given due to the
local recrossing-free property of the \ac{DS} presented here (see
Sec.~\ref{sec:definition} for further discussion on how to construct the \ac{DS} and
its properties).

Reacting particles can, in principle,
cross the \ac{DS} at some previously unknown values of the bath
coordinates and velocities in between the pre-calculated points,
and at a previously unknown time $t$
(schematically visualized as trajectory in Fig.~\ref{fig:sketch_reaction}),
while the DS itself is a time-dependently
moving object.
To obtain a rate from the simulation of a large number of trajectories,
one therefore needs
to evaluate the location $x^{\mathrm{DS}}(\vec{y}, \vec{v_y}, t)$ of the \ac{DS}
for a wide range of the parameters $\vec{y}$, $\vec{v_y}$, $t$, and the
parameter space will typically have high dimension.
However, even the
calculation of $x^{\mathrm{DS}}(\vec{y}, \vec{v_y}, t)$ at single points is
challenging and numerically expensive.
Indeed, it is hopeless to calculate
$x^{\mathrm{DS}}(\vec{y}=\vec{y}^{p}, \vec{v_y} = \vec{v_y}^{p}, t)$
from scratch for every position $(\vec{y}^{p}, \vec{v_y}^{p})$ on the trajectory
of every particle $p$ at the respective time $t$,
while propagating a large ensemble.
Consequently, it is critical to
obtain a representation of the function
$x^{\mathrm{DS}}(\vec{y}, \vec{v_y}, t)$,
that describes the dependence of the \ac{DS}
on all its arguments over a
sufficiently large range, and which is easy to evaluate.

We obtain a computationally efficient representation of the \ac{DS} in two
steps:
First, a high-accuracy algorithm is used to obtain
a small, but incomplete, number of representative points on the \ac{DS}.
In Sec.~\ref{sec:points_on_nhim}, we present an algorithm that is able
to efficiently calculate these points, based on the geometric properties of
phase-space in the saddle region.
Second, a smooth representation of the
function $x^{\mathrm{DS}}(\vec{y}, \vec{v_y}, t)$
is obtained through an efficient extrapolation of the finite
points obtained in the first step.
This requirement can sometimes be achieved through interpolation and
fitting of expected forms of the potential.
To avoid such numerics and to take advantage of the limited data
set presently available, we have recently implemented
machine learning approaches which should, in principle, be applicable
to arbitrarily high dimension \cite{hern18c}.
The use of these methods to obtain a
time-dependent description of the full \ac{DS}
is summarized, extended and assessed in Sec.~\ref{sec:machine}.

\subsection{Obtaining rate constants from reactant decay}
\label{subsec:rate_calculation}

Through \ac{TST} one aims to acquire rate constants for a chemical reaction from a reactant
state over an energy barrier to a product state.
A typical approach for obtaining rate
constants~\cite{Haenggi1990} is to propagate an ensemble of particles,
identify only those trajectories from reactants to products
that cross the \ac{DS} such as that illustrated in Fig.~\ref{fig:sketch_reaction},
sum their flux, and divide by the population of reactant particles.
In the \ac{TST} approximation the reacting flux is obtained by summing
the instantaneous flux across the entire \ac{DS}.
Rates thus obtained are exact if and only if the \ac{DS} is free of recrossings.

To obtain rate constants, we monitor the number of
trajectories that remain reactants over time
$N\sno{react}(t)$ before crossing the moving \ac{DS}.
For large times, when all
trajectories have left the barrier region, it approaches the number of
non-reactive particles $N_\infty$.
After the initial transient behavior has decayed,
this limit is approached exponentially
\begin{equation}
  N\sno{react}(t)-N_\infty \propto \exp(-k\,t)\, ,
  \label{rate_constant}
\end{equation}
where $k$ is the
rate constant of this reaction.\cite{chan78,pech81,miller83}
Deviations from exponential behavior also arise at long
times when the number of remaining trajectories becomes too small to be
statistically meaningful.

\section{Obtaining single points on the \ac{NHIM}}\label{sec:points_on_nhim}

\subsection{Revealing phase space structures}

The phase space structures discussed in
Sec.~\ref{sec:definition} can be uncovered through an increasing
array of tools, including the ones reviewed in this section.
For simplicity, we will illustrate them using
a one-dimensional model system for a simple
chemical reaction with a periodically moving barrier and open reactant and
product channels.
Specifically, we consider the motion of a particle of unit
mass under the influence of the time-dependent potential
\begin{align}
  V(x, t) &=  E\sno{b}\exp\left\{-{[x - \hat x\sin(\omega_x\,t)]}^2\right\}.
  \label{1d_model}
\end{align}
Here, $E\sno{b}$ is the height of a Gaussian barrier oscillating along the $x$
axis with frequency $\omega_x$ and amplitude $\hat x$.
We use dimensionless
units in which we set $E\sno{b} = 2$, $\hat x = 0.4$, and $\omega_x = \pi$.
\begin{figure}[t]
  \includegraphics[width=\figurewide]{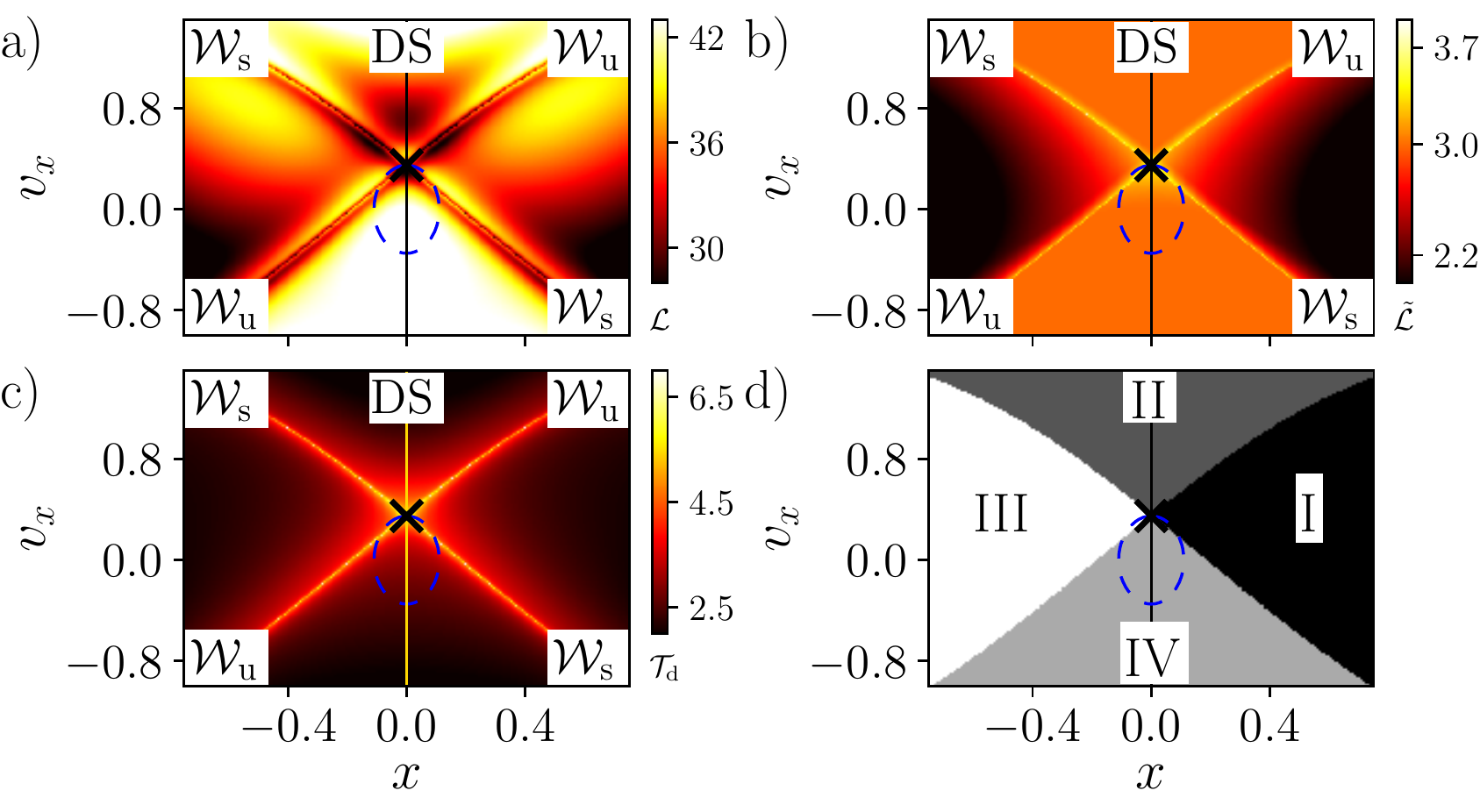}
\caption{Phase space of the system introduced in Eq.~\eqref{1d_model}.
Several
methods are visualized to reveal the stable ($\mathcal{W}\sno{s}$) and
unstable ($\mathcal{W}\sno{u}$) manifolds attached to the saddle.
As in Fig.~\ref{fig:reactive_regions}, their intersection (the
\ac{NHIM} at a given time) is marked with a black cross,
and the \ac{DS} attached
to it is a vertical solid line.
The blue dashed curve is the union of all
\acp{NHIM} points (and therefore the \ac{TS}-trajectory) for the full time-dependence
of Eq.~\eqref{1d_model}.
The contours in the panels correspond to
(a) $\mathcal{L}$ as defined in Eq.~\eqref{LD},
(b) $\tilde{\mathcal{L}}$ according to Eq.~\eqref{eq:LD_mod},
(c) the time-descriptor ${\mathcal{T}}_{\mathrm{d}}$ (see Eq.~\eqref{eq:TD}),
and (d) reactive and non-reactive regions
according to Table~\ref{areas}.
}
\label{fig:crosses}
\end{figure}

A tool to resolve the manifolds in phase space is the \ac{LD}.\cite{Mancho2013}
In the context of \ac{TST}, the \ac{LD} at position $\vec{x}_0$ in the coordinate space,
velocity $\vec{v}_0$, and time
$t_0$, is defined as~\cite{Mancho2013,hern15e,hern16a,hern16i}
\begin{equation}
\mathcal{L}(\vec{x}_0, \vec{v}_0, t_0) = \int_{t_0 - \tau}^{t_0 + \tau}
||\vec{v}(t)||\,\no{d}t \,.
\label{LD}
\end{equation}
The \ac{LD} measures the arc length of the trajectory $\vec{x}(t)$ that passes
through $(\vec x_0, \vec v_0)$ at time $t=t_0$.
The trajectory is considered
forward and backward in time over the  interval $[t_0-\tau;~t_0+\tau]$, and the
parameter $\tau$ is chosen such that the relevant time scale of the system is
covered.
In the case of open reactant and product basins, the stable
$\mathcal{W}\sno{s}$ and unstable $\mathcal{W}\sno{u}$ manifolds associated
with the energy barrier should be revealed by local minima of the \ac{LD}
in phase space, but are sometimes obfuscated by ditch-like structures, as seen in
Fig.~\ref{fig:crosses}(a).
Trajectories starting near the stable and unstable
manifolds generally have low \acp{LD} because they are bound to the saddle
region for a finite time.
One ditch ---that is, a deep minimum---
is found in the forward time component, indicating the location of the stable manifold
$\mathcal{W}\sno{s}$ and the other ditch is found in the backward time component, indicating the unstable manifold $\mathcal{W}\sno{u}$.

In the case of closed reactant and product basins, the \ac{LD} defined in
Eq.~\eqref{LD} results in a fractal-like structure in phase space that also contains
both mani\-folds, but obfuscates them even more~\cite{hern17e}.
Nevertheless, $\mathcal{W}\sno{s}$ and
$\mathcal{W}\sno{u}$ can be revealed if \acp{LD} are computed not for a fixed time $\tau$,
but rather until they leave the saddle region
$x_\text{R} \le x \le x_\text{P}$:
\begin{eqnarray}
   \tilde{\mathcal{L}}(\vec{x}_0, \vec{v}_0, t_0) &=&
 \int_{t_\text{b}(\vec{x}_0,\vec{v}_0,t_0)}^{t_\text{f}(\vec{x_0}, \vec{v_0},t_0)}
	 ||\vec{v}(t)||\,\no{d}t,
\label{eq:LD_mod}\\
 x \left(t_\text{b,f}(\vec{x}_0,\vec{v}_0, t_0)\right) &=&x_\text{R} \text{~or~} x_\text{P}, \quad\mbox{for }
   t_\text{b} < t_0 < t_\text{f} \nonumber.
\end{eqnarray}
As shown in Fig.~\ref{fig:crosses}(b), the manifolds are then clearly revealed by
local maxima in $\tilde{\mathcal{L}}$.

The shift from minima in $\mathcal{L}$ to maxima in the modified
$\tilde{\mathcal{L}}$ is attributed to the fact that trajectories close to the
stable and unstable manifold are integrated over longer times, while
the respective particles are
moving in the saddle region.
With this observation, one can find a simpler
and arguably more reliable approach to reveal the manifolds.
By simply tracking
the time a particle spends in the saddle region, we alleviate the need to
compute the arc length of the trajectory.
Following the nomenclature of the
\ac{LD}, we refer to these times as \ac{TD}:
\begin{eqnarray}
\label{eq:TD}
	\mathcal{T}_\text{d}(\vec x_0, \vec v_0, t_0) &\equiv&
		 t_\text{f}(\vec x_0, \vec v_0, t_0) - t_\text{b}(\vec x_0, \vec v_0, t_0),  \nonumber \\
	\mathcal{T}_\text{d,f}(\vec x_0, \vec v_0, t_0) &\equiv&
		 t_\text{f}(\vec x_0, \vec v_0, t_0) - t_0, \\
	\mathcal{T}_\text{d,b}(\vec x_0, \vec v_0, t_0) &\equiv&
		 t_0 - t_\text{b}(\vec x_0, \vec v_0, t_0). \nonumber
\end{eqnarray}
The closer a particle is to $\mathcal{W}\sno{s}$ in forward time or to
$\mathcal{W}\sno{u}$ in backward time, the longer it will stay in the saddle
region in the respective time component. Consequently, the stable and the unstable manifold are again revealed by
the maxima of the \ac{TD}, as seen in Fig.~\ref{fig:crosses}(c).
Note that since trajectories on the stable and unstable manifolds are trapped in
the barrier region for infinite time, both $\tilde{\mathcal{L}}$ and $\mathcal{T}_\text{d}$ may
diverge, although it is unlikely when propagating the trajectories numerically.

\begin{figure}[t]
\centering
  \includegraphics[width=0.8\figurewide]{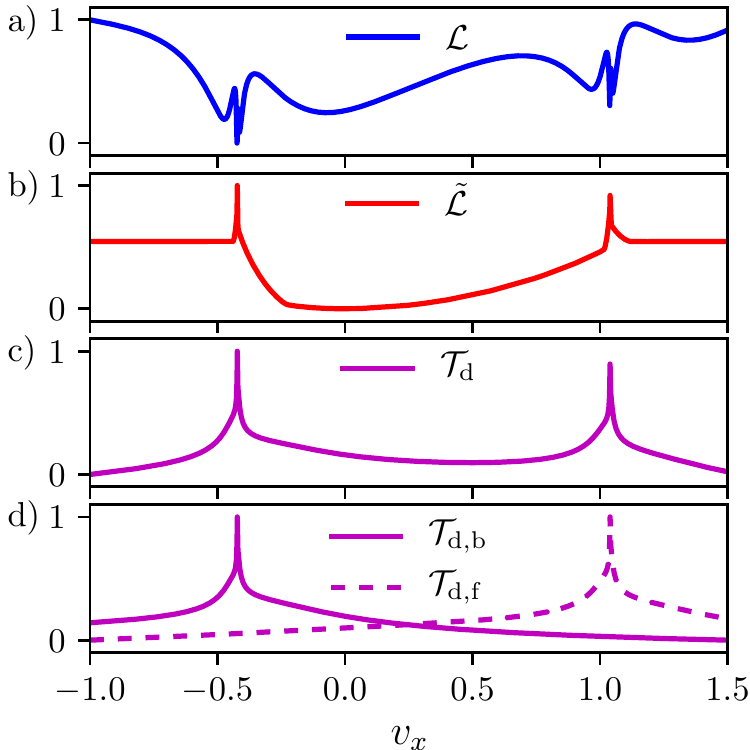}
\caption{One-dimensional view along $v_x$ of the data in Fig.~\ref{fig:crosses}
at $x = -0.4$. The \ac{LD} without cutoff is displayed in (a), the \ac{LD} with
cutoff in (b) and the \ac{TD} in (c). Subfigure (d) shows the individual parts
of the \ac{TD} for propagation solely in backward respectively forward time. All
curves have been normalized.
}
\label{fig:1d_manifolds}
\end{figure}

To calculate the $v_x$-position of the two manifolds in phase space for a given
$x$-position (reaction coordinate),
one can search for the
appropriate extremum
of the \ac{LD}, the
modified \ac{LD}, or the \ac{TD}. Using what we refer to as the standard \ac{LD}
is perhaps the least robust of
these methods because the \ac{LD} has several
local minima in addition to the primary minimum indicating the manifold,
see Fig.~\ref{fig:1d_manifolds}(a).
The modified~\ac{LD} and the \ac{TD} do not show
such substructure and have clear maxima for both stable and unstable
manifolds, see Fig.~\ref{fig:1d_manifolds}(b) -- (d).
However, because of its simpler definition the \ac{TD} seems to be
the method of choice as long as one can clearly define a saddle
region.
With all of these methods, the integration in forward and
backward time can also be performed separately to reveal the two
manifolds individually, see Fig.~\ref{fig:1d_manifolds}(d).
Thus, numerical ambiguities between stable and unstable manifolds
near their intersection can be avoided.

\subsection{1D: Finding the NHIM}\label{methods_1d_nhim}
With the stable and unstable manifold revealed by the \acp{LD} or \acp{TD}, it is now possible
to find the \ac{NHIM}.
Since the \ac{NHIM} is given by the intersection of the
stable and unstable manifold, it can be computed by
finding coordinates where their distance is zero.
This can be accomplished with a root search that finds
the value of the reaction coordinate $x$ for which
the difference between the
velocities $v_x$ of the stable and the unstable manifold is zero.
However, this
procedure is very costly as it nests an extreme value search
of the \acp{LD} or \acp{TD} within a root search for the distance
between these extreme values.
Moreover, the computation of
each of the many \ac{LD} or \ac{TD} values underlying this search
is itself expensive because it requires the
integration of a full trajectory.

\begin{figure}[t]
   \centering
  \includegraphics[width=\figurewide]{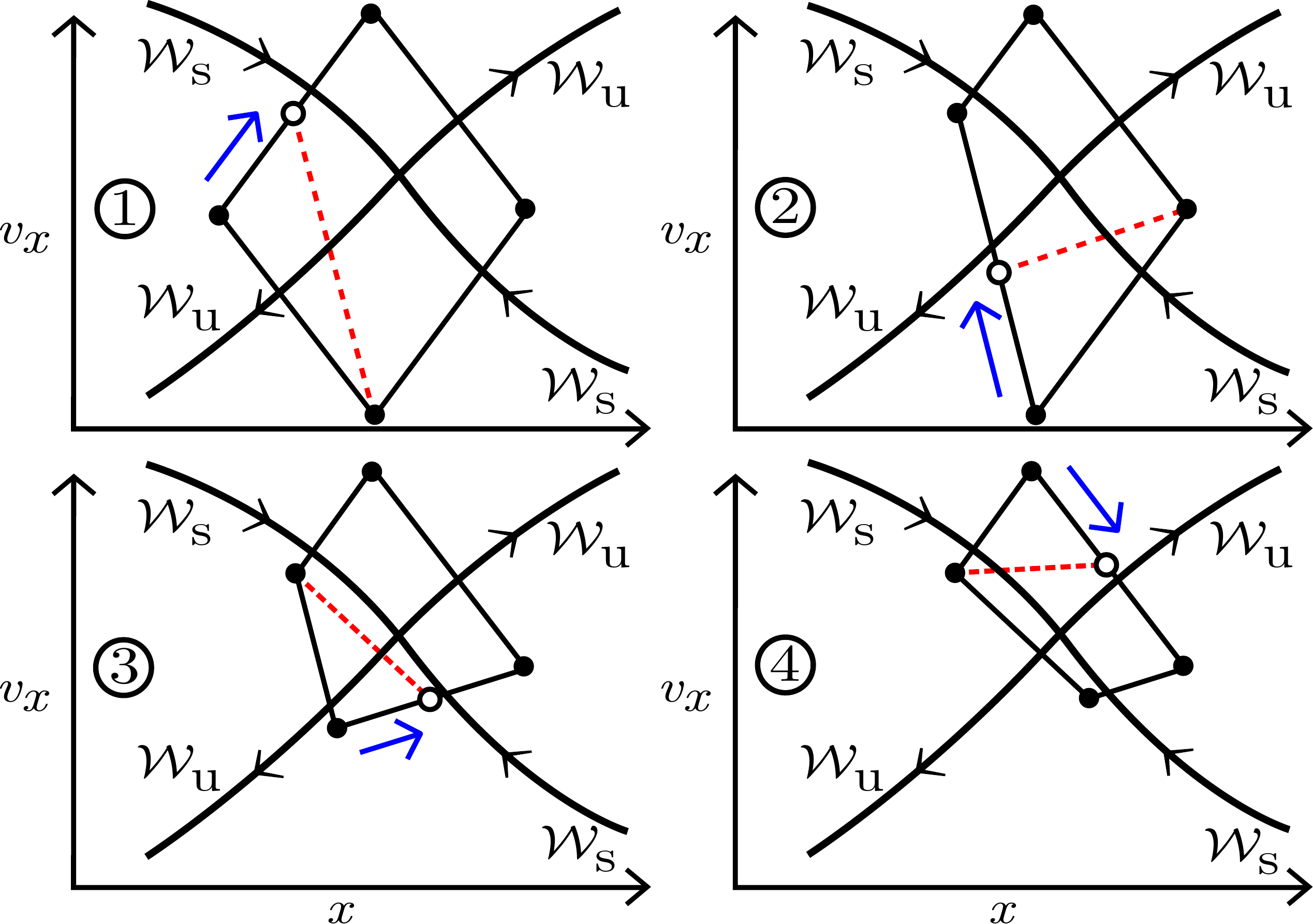}
   \caption{Illustration of the binary contraction method. The algorithm starts
     with a quadrangle having each of its vertices (black bullets) in one of the
     reactive and non-reactive areas described in Table~\ref{tab:regions}.  At
     each step, a new trajectory starting at the center between two of the
     vertices is calculated (white bullet) and classified to one of the areas.
     Afterwards the corresponding vertex of the quadrangle is replaced.
     Iteration of this procedure for each distance between adjacent
     vertices shrinks the quadrangle to the intersection of the stable
     $\mathcal{W}\sno{s}$ and unstable $\mathcal{W}\sno{u}$
     manifold.}
     \label{fig:algorithm}
\end{figure}

An efficient alternative for obtaining the \ac{NHIM}
is the binary contraction method,\cite{hern18g}
which is based on the classification of trajectories in
Fig.~\ref{fig:crosses}(d).
The algorithm starts with a quadrangle whose vertices
represent the four different regions, see Fig.~\ref{fig:algorithm}.
A candidate vertex is then associated
with the midpoint of one of the edges of the quadrangle,
and a trajectory is initiated to determine which region it
belongs to.
The corresponding vertex is then replaced, and the
area of the quadrangle is reduced.
This process is repeated for all vertices until the quadrangle
converges to the NHIM with the desired accuracy
---that is, sufficiently small area.
Additional details of the algorithm,
including an explanation on how to find good initial conditions
and the treatment of exceptional cases,
can be found in Ref.~\citenum{hern18g}.

Thus the binary
contraction method can be used to find the
\ac{NHIM} (or the \ac{TS} trajectory) with high precision,
while propagating considerably fewer trajectories
than the optimization method described above.
For example,
in the one-dimensional model potential of Eq.~\eqref{1d_model}
with periodic external driving,
this leads to a periodic orbit (the \ac{TS} trajectory), as
demonstrated by the blue dashed circles in Fig.~\ref{fig:crosses}.

\subsection{The NHIM in higher dimensions}\label{multidim_sys}

In systems with $n>1$ degrees of freedom, the situation is qualitatively
similar, but it is more challenging because all phase space structures have
higher dimensions.
As already established in Sec.~\ref{subsec:RecrossFreeDS},
the \ac{NHIM} has dimension $2n - 2$ in phase space.
Consequently, its
associated stable and unstable manifolds both have dimension $2n - 1$.
The time-dependence in the potential,
either because of the external driving or the noise,
adds additional complexity which makes these manifolds
time-dependent.

A point in phase space can be characterized by the reaction coordinates $x$ and
$v_x$ and the bath coordinates $\vec y$ and $\vec{v_y}$.
For any fixed values of
the bath coordinates $\vec y$ and $\vec{v_y}$, we obtain a two-dimensional
section through phase space with coordinates $x$ and $v_x$.
We assume that in
each of these sections the stable and unstable manifold intersect
as shown in Fig.~\ref{fig:reactive_regions}.
In particular, each section contains a single
point of the \ac{NHIM} that lies at the intersection of the two reactive and
two nonreactive regions. This condition provides a unique reaction coordinate
and velocity
$(x^{\no{NHIM}}(\vec{y},\vec{v_y}, t), v_x^{\no{NHIM}}(\vec{y}, \vec{v_y}, t))$
on the \ac{NHIM} for each bath coordinate and time, which can be found by the
binary contraction algorithm.
Because these coordinates depend on $2n-2$
variables $\vec y$ and $\vec {v_y}$ in phase space,
the \ac{NHIM} has dimension $2n-2$.

A possible counterexample arises in
the simple limiting case of an uncoupled harmonic system.
Therein, the point in the reactive phase space,
$x^{\mathrm{NHIM}}$ and $v_x^{\mathrm{NHIM}}$
will be constant, independent of the bath coordinates $\vec y$
and $\vec {v_y}$.
In this case, it would not be possible to represent any of the bath modes
as functions of $x$ or $v_x$.
Nevertheless, the construction above provides the correct phase
space point because it does not rely on the association being
injective.

A non-trivial multidimensional example is a
two-dimensional model with a
time-periodically moving barrier
separating open reactant and product
basins.
It can be represented using an extension of
the one-dimensional potential of
Eq.~\eqref{1d_model} to
\begin{eqnarray}
V(x, y, t) &=&
E\sno{b}\,\exp\left(
  -{\left[x- \hat{x} \sin\left(\omega_x t \right)\right]}^2\right) \nonumber \\
&&+ \frac{\omega\sno{y}^2}{2}{\left[y-\frac{2}{\pi}
\arctan\left(2 x \right)\right]}^2.
\label{eq:potential}
\end{eqnarray}
This potential describes a two-dimensional potential energy
landscape that has minima in the bath coordinate along the curve $y=({2}/{\pi})
\arctan\left(2 x \right)$.
A moving Gaussian barrier with height $E\sno{b}$ is
added over this stationary potential.
It is oscillating along the $x$
axis with frequency $\omega_x$ and amplitude $\hat x$.
$\omega_y$ is the
frequency of oscillations in the bath mode for a particle of unit mass.
For simplicity, we again
use dimensionless units in which the parameter are $E\sno{b}=2$, $\omega_x=\pi$,
$\omega\sno{y}=2$, and $\hat x = 0.4$.

\begin{figure}[t]
  \includegraphics[width=0.8\figurewide]{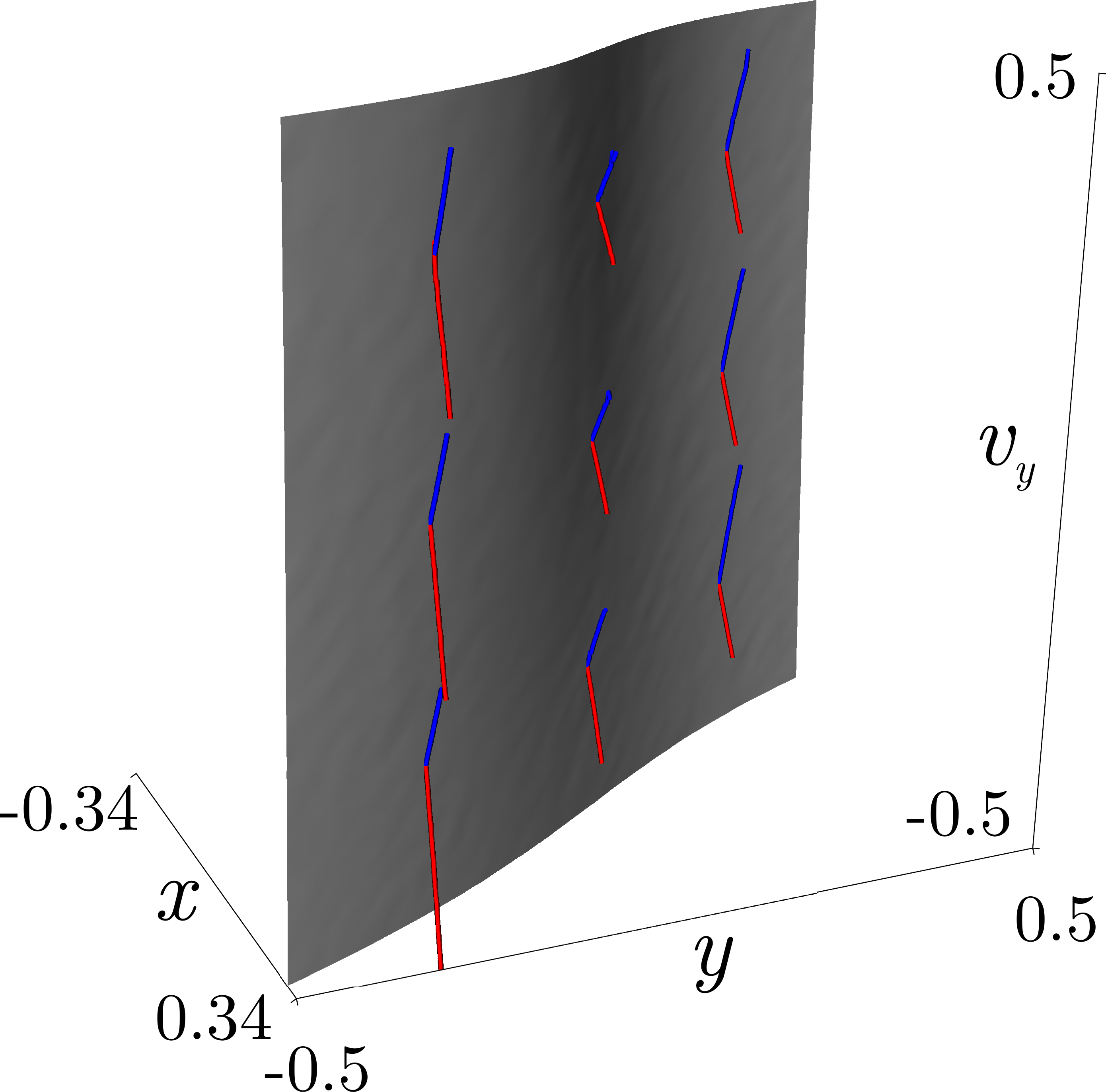}
   \caption{Gray surface: $x^{\mathrm{NHIM}}(y, v_y)$ for the
     potential~\eqref{eq:potential} at $t=0.5$ with $y$
     and $v_y$ varied on an equidistant grid in the plotted range.
     For $(x, v_y)$-slices at $3\times 3$ different values $(y, v_x)$
     the one-dimensional fiber of the stable manifold in each slice
     is given as a red line, the fiber of the unstable manifold as a blue line.
     Their intersections are located on the two-dimensional \ac{NHIM}.
}\label{fig:crosses_with_surface}
\end{figure}

We choose $x$ as the approximate reaction coordinate and $y$ as the bath
coordinate.
Consequently, we  calculate an $(x, v_x)$-slice for each set of the
remaining coordinates $(y, v_y)$ and for any time.
Using the binary contraction method,
we find the intersection ${(x, v_{x})}^{\mathrm{NHIM}}(y, v_y, t)$.
Repeating this procedure for different values $(y, v_y)$ on an equidistant grid
for any time $t$ and using an appropriate interpolation method (see below), we
obtain the full \ac{NHIM} of the system.
It is a two-dimensional, time-dependently
moving surface embedded in the four-dimensional phase space. A
representation of this \ac{NHIM} at $t=0.5$ is shown in
Fig.~\ref{fig:crosses_with_surface} as a gray surface.

For certain points $(y, v_y)$, Fig.~\ref{fig:crosses_with_surface} shows
sections of the stable and unstable manifolds that are attached to the
\ac{NHIM}.
To obtain them, we consider a two-dimensional section of phase
space with fixed values of $y$ and $v_x=v_x^{\mathrm{NHIM}}(y,v_y)$.
The dynamical structure of this section is similar to that in
Fig.~\ref{fig:reactive_regions}, except that the axes are now $x$ and $v_y$.
The manifolds are obtained by finding the value of $v_y$ at which,
for given $x$,
the \ac{TD} is maximal.
The stable and unstable manifolds intersect on the
\ac{NHIM}, as expected.
Because they were obtained independently of the
calculation that yields the \ac{NHIM}, this observation
also serves to confirm the reliability of the computation
and the uniqueness of the \ac{NHIM}.

\subsection{TS-trajectory in higher dimensions}

In a one-dimensional system, the \ac{NHIM} is a single trajectory that can be
identified as the \ac{TS} trajectory.
A higher-dimensional \ac{NHIM}, by
contrast, is made up of infinitely many trajectories over time, all of which
are trapped in the saddle region for all times.
Thus a unique identification of the \ac{TS} trajectory requires
additional constraints beyond the minimal requirement that it
is trapped within the saddle region.

In an autonomous system,
the location of the DS is
often marked by
the naive \ac{TS} ---that is the saddle
at the top of the barrier.

The \ac{TS} is the point on some selected reaction
path that crosses the \ac{DS} and it is not necessarily the
same as the naive \ac{TS} but is often in its proximity.
We further note that there exist
exceptions to cases in which the \ac{TS} is
near the naive \ac{TS}.
This includes the \ac{TS} structures far from the naive \ac{TS}
found by Gray and Davis\cite{Davis-1986}
for the autonomous HeI$_2$ system
and, more recently, roaming reactions that are associated to a
\ac{TS}
that is not necessarily
close to a
saddle.\cite{bowman04a, hern13e, wiggins14a, bowman2011c,bowman14a}
Such cases are not addressed in the present work.

The \ac{TS} trajectory is thus intended to generalize
the fixed \ac{TS} for driven systems.
In a one-dimensional, time-periodically driven system, the
\ac{TS} trajectory is a periodic orbit.
Following this
property, we suggest that for a time-periodically driven, multidimensional
Hamiltonian system the period-1 orbit trapped in the barrier region,
if uniquely identified, should also be called the
\ac{TS} trajectory. Here, period-1 qualifies the orbit
as one which is periodic and for which the ratio of its
period to that of the external driving is 1.

To make this definition precise, we determine the
Euclidean distance $\Delta \gamma$ of two phase space coordinates $\vec{\gamma}
= (\vec{x}, \vec{v})^\mathrm{T}$ on the same trajectory, one time period
$T$ of the potential apart.
In practice, it can be computed by propagating
initial conditions given for a certain time $t_0$ forward and backward in time
by half a potential period $T/2$. We perform this computation in such a way to
avoid errors accumulated by propagating trajectories near unstable manifolds
over long times.
A trajectory with $\Delta \gamma = \left |
\vec{\gamma}(t_0 + T/2) - \vec{\gamma}(t_0 - T/2)\right| = 0$ is a
 periodic
orbit.
We identify such a periodic orbit as the \ac{TS} trajectory if it
is the only period-1 orbit bound to the saddle region
with respect to the periodicity of the external driving in a time-periodically
driven system.

Again, the same assumptions as discussed in Sec.~\ref{subsec:RecrossFreeDS} are valid,
meaning that the system should be not too non-linear, and the velocities of
reacting particles not too high. Such high non-linearities lead in general to
a break-down of our assumptions as the motion on the NHIM of systems with
many degrees of freedom
becomes highly chaotic. Examples for such chaotic regimes can be found
in the high-energy
limit of autonomous systems described
in Refs.~\citenum{Komatsuzaki99a,komatsuzaki06b}.
In such highly non-linear systems,
the NHIM and the TS can even bifurcate, as found in Refs.~\citenum{Li09,komatsuzaki06a}.

\begin{figure}[t]
  \includegraphics[width=0.6\figurewide]{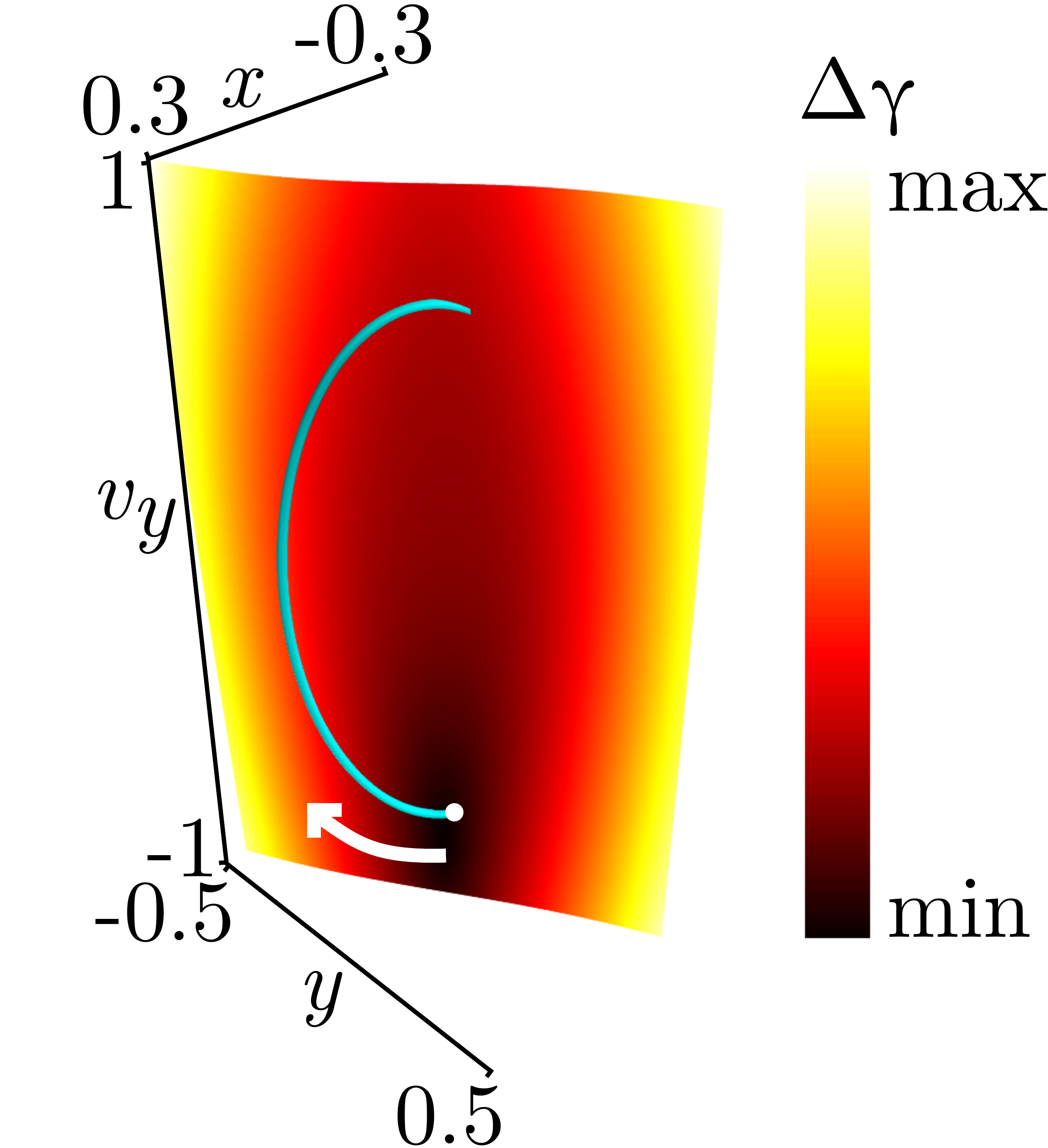}
\caption{%
  Surface: visualization of the \ac{NHIM} for potential~\eqref{eq:potential} at
  $t=0$. Its color indicates the Euclidean distance of the phase space vectors
  $\Delta \gamma = \left|\vec{\gamma}(t_0 + T/2) - \vec{\gamma}(t_0 -
  T/2)\right|$ for trajectories started at $t_0 =0$ at the respective point on
  the \ac{NHIM} and propagated for half the period time of the external driving
  in potential~\eqref{eq:potential} in forward and backward time. The point with
  $\Delta \gamma =0$ is marked by a white dot. It is the \ac{TS} trajectory at time
  $t = t_0$. Propagating a trajectory starting at this specific point yields the
  \ac{TS} trajectory displayed as a blue line (partially covered
  by the \ac{NHIM}). The direction of movement on the \ac{TS} trajectory is
  indicated by a white arrow. When calculated for different initial times $t_0$
  in a full period $T$, the union of all minima of $\Delta \gamma$ on each
  \ac{NHIM} at time $t_0$ corresponds to the \ac{TS} trajectory.
}
\label{fig:ts_traj}
\end{figure}

The distance $\Delta \gamma$ shown in Fig.~\ref{fig:ts_traj}
has been computed on the \ac{NHIM} at $t_0 = 0$ for
the potential~\eqref{eq:potential}.
The color of any point on the \ac{NHIM}
indicates the distance $\Delta \gamma$ of
the phase space vectors of two trajectories, integrated for $T/2 = 1$ in forward
and backward time.
A pronounced minimum, and thus by our proposed definition, a
point on the \ac{TS} trajectory, can clearly be distinguished.
Using this point as the initial condition for a trajectory propagation,
the periodic \ac{TS} trajectory can be obtained.
Since this propagation will fail after a certain time
because the \ac{NHIM} is unstable, one cannot compute a fully periodic
trajectory numerically.
Nevertheless, with sufficiently accurate initial
conditions on the \ac{NHIM}, one can compute a periodic trajectory up to a
certain error tolerance.
Alternatively, various points on the \ac{TS} trajectory
can be computed from scratch searching for the minimum of $\Delta \gamma$ on the
\ac{NHIM} calculated at different initial times $t_0$.

\section{Machine-learning based description of the \ac{NHIM}}\label{sec:machine}

Using the methods described in Sec.~\ref{sec:points_on_nhim}, a single point on the
\ac{NHIM} can be calculated for arbitrary values of the bath modes and 
time\EDITS{, as is done here for the
illustrative case of system \eqref{eq:potential}}.
However, for a rate calculation, the position of the \ac{DS} must be compared to
the instantaneous position of each propagated particle at each time step.
To compute the required reaction coordinate of the \ac{DS}
from scratch whenever it is
needed would be prohibitively expensive because even the efficient binary
contraction method described above
requires the propagation of dozens of trajectories.
To reduce the numerical effort,
it is critical to obtain a continuous representation of the
\ac{DS} as a function of bath modes and time that is based on the knowledge of
a comparatively small number of points on the \ac{NHIM}.

Machine learning methods have
already been applied within the field of theoretical chemistry
for the interpolation
of high-dimensional potential energy surfaces.\cite{blank1995neural,behler2007generalized,behler2011atom,cui2016efficient,vargas2017machine,
rupp2012fast,cui2015gaussian_a,cui2015gaussian_b,faber2016machine,huang2017chemical,carrington06}
Following our recent work,\cite{hern18c}
we use machine learning techniques
---and specifically through the use of \acp{NN} or
\ac{GPR}---
to characterize an entirely different set of surfaces:
the \ac{NHIM} and the \ac{DS} attached to it.
After training on an arbitrary set of
points on the \ac{NHIM}, a neural network yields a continuous representation
that approximates the \ac{NHIM} at the given points and elsewhere as shown
in Sec.~\ref{neural_networks}.
In principle,
the method can be applied for arbitrarily high dimension, though the complexity
of the network may increase as the number of bath coordinates grows.
Alternatively, the interpolation of a given set of points on the \ac{NHIM}
can be performed using
\ac{GPR} (see, e.\,g.~Ref.~\citenum{murphy2012machine} and
references therein)
as discussed in Sec.~\ref{kernel_methods}.
This method requires \emph{a priori} knowledge of the length and
time scales along which the position of the \ac{NHIM} is expected to vary.
The comparison between these two methods and a discussion of their advantages
and disadvantages in a given context will follow in Sec.~\ref{comparison}.

\subsection{Feed-forward neural networks}\label{neural_networks}

Neural networks (\acp{NN})
are a powerful tool for approximating functions of
arbitrary complexity in any dimension.\cite{FUNAHASHI1989183}
Inspired by our current understanding of \acp{NN} in the brain, an
artificial \ac{NN} consists of a set of model neurons with complex
connections between them.
It transforms a given input into some output.
Typically, a \ac{NN} is trained in the sense that
a set of free parameters is adjusted
on a set of input points for which the output is known.
During \emph{training},
the net \emph{learns} to find patterns and dependencies in the given data
without requiring any user input beyond the training data.
When properly
trained, the network will provide a good approximation to the correct output for
input similar to the points it was trained on.
In our case the training data
consists of a set of points on the \ac{NHIM}.
By training on these points, we
can create two separate \acp{NN} that represent the functions
$x^{\mathrm{NHIM}}(\vec{y}, \vec{v_y}, t)$ and $v_x^{\mathrm{NHIM}}(\vec{y},
\vec{v_y}, t)$ separately, or use a more complex network to represent the
combined function $(x, v_x)^{\mathrm{NHIM}}(\vec{y}, \vec{v_y}, t)$.
This representation will be reliable in the range of inputs that a suitably chosen
set of training data covers, as long as the accuracy of the output was verified.

\begin{figure}[t]
  \centering
  \includegraphics[width=1.0\figurewide]{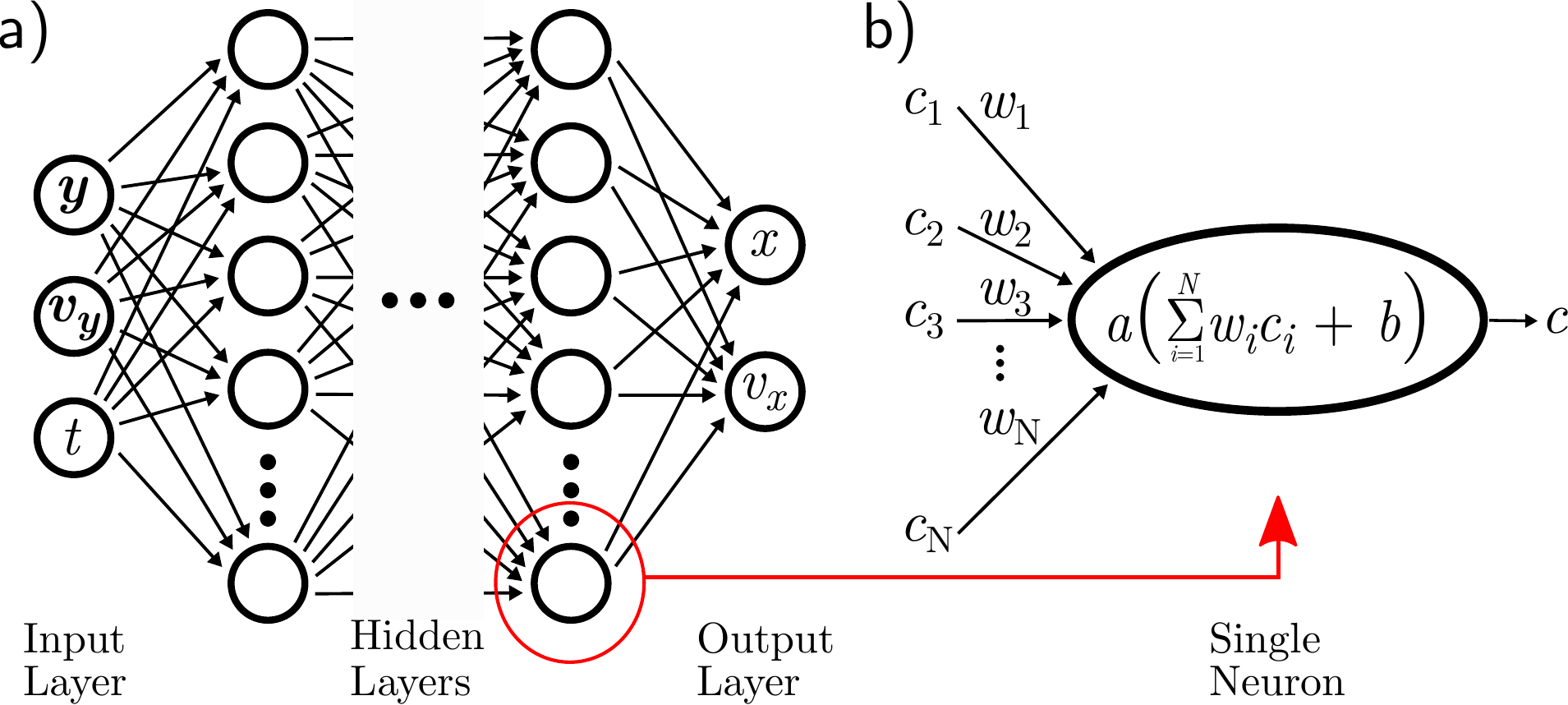}
\caption{%
  (a): A schematic for a \ac{NN} approximating the function ${(x,
  v_{x})}^{\mathrm{NHIM}}(\vec{y}, \vec{v_y}, t)$. Using the $(n-1)$ bath
  coordinates $\vec{y}$, the respective velocities $\vec{v_y}$ as well as a time
  $t$ as input, the net processes this information layer by layer, passing the
  activation signals of the neurons from the previous layer to the next. The
  arbitrary number of layers between input layer and output layer are the hidden
  layers. Each layer has a number of neurons which interpret the incoming
  signals and pass them along to eventually calculate the reaction coordinates
  ${(x, v_x)}^{\mathrm{NHIM}}$ as results of the output layers.
 (b): A schematic of a
  neuron, indicating how its activation is computed. It sums up the input
  signals from the previous layer of neurons $c_i$ weighted by $w_i$ and adds an
  associated bias $b$, passing the results into the activation function
  $a(x)$ to compute its output $c$.
}
\label{fig:neural_net}
\end{figure}

Here, we provide a brief overview on how we set up and use
a \ac{NN} to represent the \ac{DS}.\cite{hern18c}
For a more general discussion on \acp{NN}, we refer
the interested reader to the literature,
e.\,g.~Refs.~\citenum{Goodfellow-et-al-2016,Nielsen-2015} and
references therein.
A feed-forward \ac{NN} consists of several layers
through which information is processed, see illustration in the schematic of
Fig.~\ref{fig:neural_net}.
As we create a net to approximate ${(x,
v_{x})}^{\mathrm{NHIM}}(\vec{y}, \vec{v_y}, t)$, the input layer is given by the
bath coordinates $\vec{y}$, velocities $\vec{v_y}$, and time $t$.
Based on
its input, each neuron computes an output value that is passed to the next layer
of the network, from that layer to the next, and so on until the final output is
obtained.

In typical neural network applications, a user must decide on the number of
hidden layers, the number of neurons in each layer, and the function by which a
neuron converts its input into its output.
Each neuron ---such as the one shown in Fig.~\ref{fig:neural_net}(b)---
of a given layer
receives inputs $c_i$ from
the~$N$ neurons of the previous layer.
Its output,
\begin{equation}
  c = a\! \left(\sum_{i=1}^{N} w_i\,c_i + b\right)\,,
  \label{activation_function}
\end{equation}
depends on the input values and on the activation function $a(x)$
whose nonlinearity ensures that the \ac{NN} is not simply
a linear transformation of the input variables and hence capable of
capturing the necessary complexity.
In this work, we choose the activation function $a(x) = \tanh\left( x\right)$
for all neurons except those on the output layer
as it is smooth and bound for $x\to\pm\infty$ while providing the requisite
nonlinearity.
The different weights $w_i$ that the neuron assigns to the output of
each neuron of the previous layer and the bias~$b$ are
optimized individually during the training process.
\EDITS{We note, that although a single hidden layer NN is a universal approximator,
it may not be the most efficient NN to a specific problem.
In the following, we tried several different NN configurations,
varying the number of hidden layers and neurons.
The examples presented below provide the best results among the
different approaches we sampled, but may not necessarily be fully
optimized.}

At the onset of training, a \ac{NN} is initialized with
the weights and biases of the neurons
chosen randomly within a reasonable range of values.
After successful training, a \ac{NN} should return a good approximation
of the outputs over the domain of inputs.
A \emph{cost function} is selected to
measure the quality of the approximation.
In the present application,
we average the mean square difference between the output of
the network and the known values over all training points.
\EDITS{This mean square cost function has units of
length squared whose values are denoted with respect to the dimensionless
units introduced in Sec.~\ref{multidim_sys}}.
To minimize the cost function,
there are several optimization routines that adjust the weights and biases of
the neurons iteratively.
This includes the \emph{back propagation algorithm}
which is a
gradient-based method that successively adjusts all free
parameters.\cite{williams1986learning,hern18c}
Usually, a \ac{NN} is exposed to
the entire set of training data many times before it
is optimized,
and each such exposure is called an \emph{epoch}.

The training of a \ac{NN} is usually not performed
with all the available data.
Instead, the data set is split into two
disjoint sets: a \emph{training set} and a \emph{verification set}.
The \ac{NN} is trained only on the training set.
The cost function on the verification set
is used to assess the reliability of the \ac{NN}
for data that it was not trained for.
If the cost function on the training set is significantly smaller than the cost
function on the verification set, then the \ac{NN} adapted very specifically to 
the given training set.
This kind of overfitting is, in general, an
undesirable effect as it indicates that the \ac{NN} is no longer learning to
make more accurate predictions.
\begin{figure}[t]
  \centering
  \includegraphics[width=0.9\figurewide]{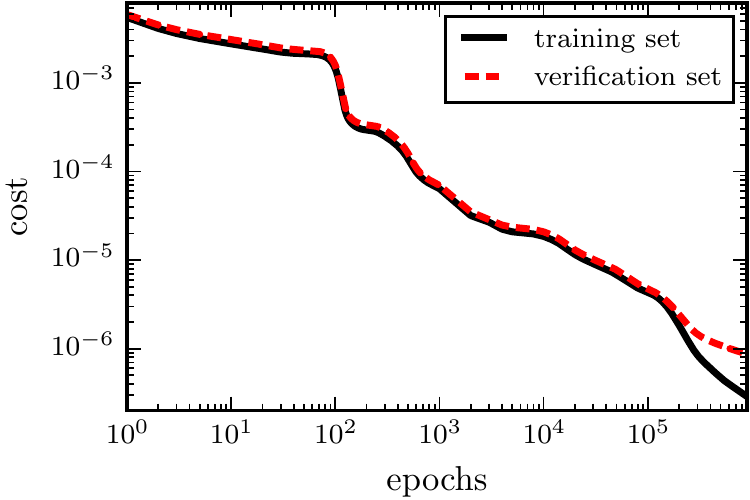}
\caption{%
Training and verification costs of a \ac{NN} for different epochs of
training.
The training is completed when the verification cost ceases to
decrease, thereby preventing overfitting to the training set.
}\label{fig:cost_function}
\end{figure}

To illustrate the ease in training a \ac{NN} to approximate the
position $x^{\mathrm{NHIM}}$ for the \ac{NHIM} in the two-dimensional model
system of Eq.~\eqref{eq:potential},
we wrote and tested a simple C++ code
without using the more generally available libraries that we
use in the production runs described below.
The network was constructed with two hidden layers:
the first and second with~$40$
and $10$ neurons, respectively.
It was trained on 2000 points
randomly distributed over the NHIM and
obtained using the binary contraction method.
The remaining data points ---numbering approximately 8000---
were used for verification.
Fig.~\ref{fig:cost_function} shows
the training and verification costs
as a function of the epochs in the training.
During training, both costs decrease monotonically at varying
speeds and in concert with each other for most of the epochs.
After more than about $10^5$ epochs,
the verification cost ceases to decrease
while the training cost decays further.
It is at this point (when additional training ceases to improve the verification)
that the training of a \ac{NN} should cease.
Additional training simply provides better agreement with the training data
while possibly distorting the general accuracy of the \ac{NN}.
In this example, we see that significant accuracy of the NHIM can be obtained
(down to a cost \STRIKE{or relative error} of $10^{-6}$, \EDITS{measured in
units of dimensionless length squared as described in Sec.~\ref{multidim_sys}}),
with a relatively small \ac{NN}.

\subsection{Gaussian process regression}\label{kernel_methods}

\acf{GPR} offers an alternative to the NN
as a supervised machine learning approach for
fitting a multidimensional hypersurface such as the \ac{NHIM} to
isolated training points.
This approach interpolates given values of a function $f(\vec z)$ that depends
on an arbitrary number of input variables~$\vec z$. In our application, the
input variables are the bath coordinates $\vec y$, velocities $\vec {v_y}$ and time~$t$.
The unknown functions ${x}^{\mathrm{NHIM}}(\vec{y}, \vec{v_y}, t)$
and~${v_x}^{\mathrm{NHIM}}(\vec{y}, \vec{v_y}, t)$ will be interpolated
independently.

\ac{GPR} is based on the assumption that the values of a function
$f(\vec z)$ are Gaussian random variables that are \emph{a priori} unknown, apart from
their statistical properties. After some values of the function have been
\emph{learned}, other values are inferred from the given information. For this
inference to be possible, some \emph{a priori} knowledge is required
about the length scales along which the values of the function vary.
Specifically, we assume that we know the \emph{a priori} mean $\mu(\vec z)$ of
the function~$f(\vec z)$ and the covariance $k(\vec z_1,\vec z_2)$ between the
values $f(\vec z_1)$ and $f(\vec z_2)$.
In an unbiased {\it a priori} distribution these means would be set to zero,
{\it i.e.}, $\mu(\vec z)=0$.
A typical choice
for $k(\vec z_1,\vec z_2)$ is the \emph{squared exponential} kernel
\begin{equation}\label{squared_exponential_kernel}
  k(\vec{z}_1, \vec{z}_2)
	 = \exp \left(-\frac{\left | \vec{z}_1 - \vec{z}_2 \right |^2}{2l^2}\right),
\end{equation}
where the \emph{hyper-parameter}~$l$ gives the typical length scale along which the
values of $f$ are expected to vary. It yields high covariance between closely
located points $\vec{z}_1$ and $\vec{z}_2$ and a lower covariance between
distant points. In an application to \ac{TST}, the characteristic length~$l$
could differ for time and space coordinates, or even for different
bath modes.

Assume now that the values
$\vec f_\text{train} = {\left(f(\vec z^{(1)}_\text{train}), \dots, f(\vec z^{(n)}_\text{train}) \right)}^\text{T}$
of the function $f$ are known in $n$ points $\vec z^{(i)}_\text{train}$
and the values
$\vec f_\text{test} = {\left(f(\vec z^{(1)}_\text{test}), \dots, f(\vec z^{(m)}_\text{test}) \right)}^\text{T}$
in $m$ other points $\vec z^{(j)}_\text{test}$ are sought.
Initially, the values $\vec f_\text{train}$ and $\vec f_\text{test}$
 follow a multidimensional Gaussian distribution with zero mean
 and covariance matrix
\begin{equation}
	\Sigma=\begin{pmatrix}
		K & N^\text{T} \\ N & M
	\end{pmatrix}
\end{equation}
 that consists of the blocks
 \begin{eqnarray}
	K_{ij} &=& k(z^{(i)}_\text{train}, z^{(j)}_\text{train})
\quad \text{(of size $n\times n$)}, \\
	M_{ij} &=& k(z^{(i)}_\text{test}, z^{(j)}_\text{test})
\quad \text{(of size $m\times m$)}, \\
	N_{ij} &=& k(z^{(i)}_\text{test}, z^{(j)}_\text{train})
\quad \text{(of size $m\times n$)}.
 \end{eqnarray}
The \emph{a posteriori}
conditional distribution of the unknown values $\vec f_\text{test}$,
 given the known values $\vec f_\text{train}$,
 is a Gaussian function with mean~\cite{murphy2012machine,rasmussen2006gaussian}
 \begin{equation}
	\label{gp_mu}
	\bar {\vec\mu} = N K^{-1} \vec f_\text{train}
\end{equation}
and covariance matrix
\begin{equation}
	\label{gp_sigma}
	\bar \Sigma = M - N K^{-1} N^\text{T}.
\end{equation}
The components of the conditional mean~$\bar{\vec \mu}$ are taken to be the
interpolated values of the function. If desired, the diagonal elements of the
conditional covariance matrix~$\bar \Sigma$ can be used to provide an error
estimate for the interpolation.
If the standard deviation
$\sqrt{\bar\Sigma_{ii}}$ is sufficiently small, the interpolated value $f(\vec
z^{(i)}_\text{test})$ is known essentially with certainty.

In practice, we compute a Cholesky decomposition $K=L L^\text{T}$ of the
matrix~$K$, where~$L$ is a lower triangular matrix.
It is then easy to pre-compute the vector
$
	K^{-1} \vec f_\text{train} = L^{-\text T} L^{-1} \vec f_\text{train}
$
that depends only on the training points, but not on the points $\vec
z^{(i)}_\text{test}$. After that, the computation of one interpolated function
value according to Eq.~\eqref{gp_mu} requires the computation of the
corresponding row of the matrix~$N$ and the scalar product with the precomputed
vector. The effort required is proportional to the number~$n$ of training
points.
The conditional covariance matrix~\eqref{gp_sigma} can be written as
\begin{equation}
	\Sigma = M - \tilde N \tilde N^\text{T},
\end{equation}
where the matrix~$\tilde N$ is the solution of the equation $N=\tilde N L^\text{T}$.
Because $L$ is triangular, the computation of one row of $\tilde N$, and
therefore of an error estimate for one data point, requires an effort of order
$n^2$.

\begin{figure}[t]
  \centering
  \includegraphics[width=1.0\figurewide]{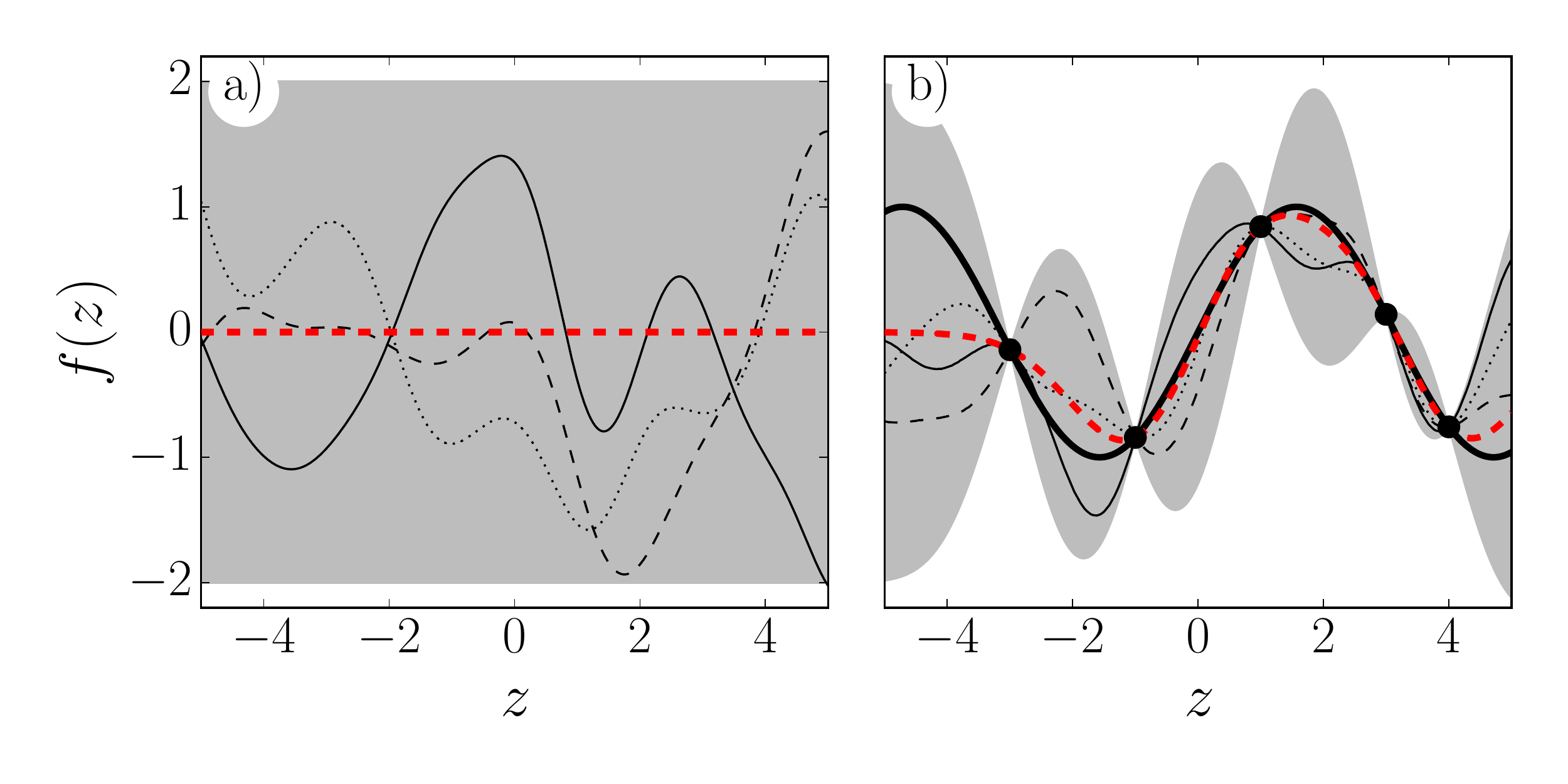}
\caption{%
  (a) A Gaussian prior with a mean of zero (thick red dashed line). The $95\,\%$
  confidence interval is given by a gray area which covers two standard
  deviations from the mean. The lines show three random functions drawn from
  this prior. (b) Five points ($\bullet$) located on a sine curve (thick black
  line) are used as training data. Learning these points turns the prior into
  the posterior distribution, changing its mean (thick red line) and its
  $95\,\%$ confidence interval.
Thin curves show three random functions sampled
  from the posterior.
  }
  \label{fig:gaussian_processing}
\end{figure}

The training and use of a \ac{GPR} machine
can be illustrated relative to
a sine function shown in Fig.~\ref{fig:gaussian_processing}~(b).
We use the squared exponential kernel~\eqref{squared_exponential_kernel} with a
characteristic length scale $l=1$. The conditional mean that serves as the
interpolated function is indicated by the dashed red line, a 95\,\% confidence
interval based on the conditional standard deviation by the gray area.

The difference between the \ac{GPR} machine's approximation
and the original sine-function is small near
the training points and increases elsewhere.
Whereas for the prior, the mean
$\mu$ and the $95\,\%$ confidence region are independent of $z$, the
\emph{learning} of several training points alters both. At the training points,
this $95\,\%$ region is zero, as these points are assumed to be correct. The
further away from training points, the more the confidence region grows.
This increase happens on a length-scale compared to the characteristic length $l=1$
of the assumed squared exponential kernel~\eqref{squared_exponential_kernel}.
At points far from
where no training points are given, e.\,g.~for $z < -3$, the confidence region
of the posterior realigns with the Gaussian prior. As a result, the accuracy of
the approximation increases with the number of training points.

To illustrate that the posterior of the \ac{GPR} machine gives
a distribution over admissible functions,
we plot three
functions randomly sampled from the posterior distribution in
Fig.~\ref{fig:gaussian_processing}~(b) and compare to three samples of the prior
distribution
in Fig.~\ref{fig:gaussian_processing}~(a).
The samples of the posterior distribution match
the training points and are mostly located within the altered $95\,\%$
confidence region around the approximation $\mu(z)$ to the original function
$f(z)$.
Therefore, the posterior `chooses' which functions of the prior are
suitable to interpolate between the given training points.

\subsection{\acp{NN} vs.~\ac{GPR}}\label{comparison}

We now compare the advantages and disadvantages of the two machine learning
methods introduced in Secs.~\ref{neural_networks} and~\ref{kernel_methods} for
the interpolation of the \ac{NHIM} and construction of the \ac{DS} in the
calculation of rate constants. The aim is to provide a guideline on when to use
which method.

For the training of a \ac{NN} on a given set of pre-calculated points of the
\ac{NHIM}, little previous knowledge on the interpolated object is needed.
For example, in a typical \ac{NN} that approximates ${(x, v_x)}^{\no{NHIM}}(\vec{y},
\vec{v_y}, t)$, the number of neurons in
each of the input and output layers are fixed
as illustrated in Fig.~\ref{fig:neural_net}.
The only information that is prescribed is the
structure of the hidden layers, and it merely needs to
contain as much complexity as the system that it is re-generating.
Its efficacy is confirmed if the optimization of the neurons
leads to weights and biases that are sufficiently small compared
to the target accuracy with respect to the cost function.

When using \ac{GPR}, the success and the speed of training highly
depends on the chosen hyper-parameters of the kernel function. In choosing these
parameters, one can include previous knowledge of the object to be interpolated,
e.\,g.~the expectation that the \ac{NHIM} of a given system may be a smooth
surface with small curvature. Consequently, the characteristic length in
the direction of the bath coordinates can be higher than the characteristic length
of the variation in time, if the external driving leads to a fast moving
\ac{NHIM} whose curvature does not vary heavily.
If such previous knowledge is available,
training can be accelerated significantly by choosing an
appropriate kernel function or
by a corresponding restriction of the parameter ranges in the optimization.

Another difference between both methods lies in the procedure of
\emph{training} itself.
For \acp{NN}, training with the back propagation
algorithm is an iterative process.
It iterates a possibly high number of times ---that is, epochs---
over the given training set gradually adjusting weights and biases to achieve a
certain level of convergence in the training and verification cost. This
procedure typically takes a lot of computation time. Using \ac{GPR},
training can be effectively broken down to basic routines in linear algebra
without need for iteration.
Therefore, training to a certain accuracy in the context of
\ac{GPR} is usually much faster compared to the use of \acp{NN}.

\begin{figure}[t]
  \centering
  \includegraphics[width=1.0\figurewide]{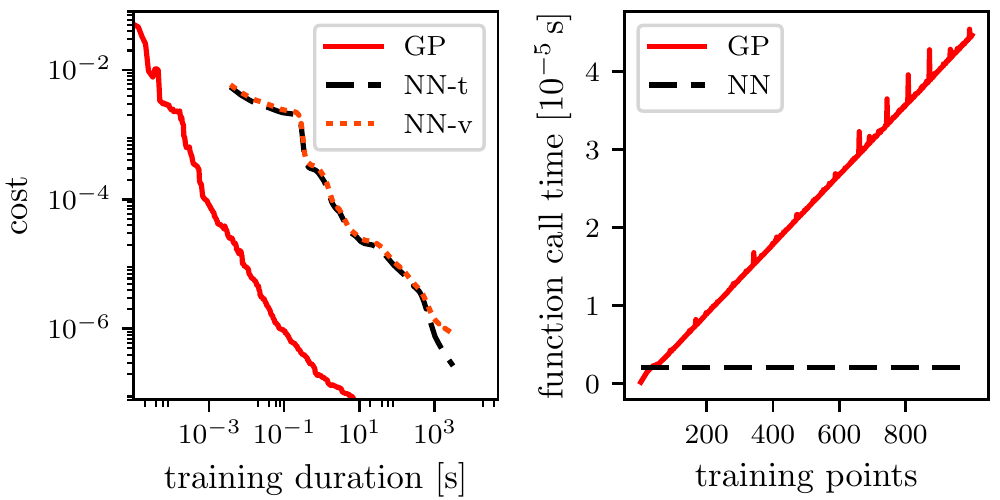}
\caption{%
  Comparison between \ac{NN} and \ac{GPR}:
In the left panel, the
  cost function over the needed training time is shown
for each implementation as a function of computing time. For \ac{GPR},
  successfully adding training points leads to a fast decrease of the training
  cost, whereas the same decrease using \acp{NN} is considerably slower
  while training with the back propagation algorithm. Here, both the training
  cost (NN-t), as well as the verification cost (NN-v) are shown.
In the right panel,
  the function call time required for each method is shown relative to the
number of training points.
For \ac{GPR}, function call time increases linearly with
  the number training points provided to it,
and which are needed to decrease the cost function.
Thus, high accuracy of
  \ac{GPR} comes with the disadvantage of an increasing function call
  time.
The black dashed line shows the function call time of a \ac{NN}
  and it is independent of the number of training points because the cost
of sampling across an epoch is dominated by the inversion of the \ac{NN}
parameters.
}\label{fig:gp_nn_comparison}
\end{figure}

We compare the training of the neural network already
discussed in Sec.~\ref{neural_networks} to the training done by
\ac{GPR} on the \ac{NHIM} of the same two-dimensional model potential
according to Eq.~\eqref{eq:potential}.
The \ac{NN} is set up with two hidden
layers with $40$ and $10$ neurons as before.
For \ac{GPR}, we use the
\emph{squared exponential} kernel (Eq.~\eqref{squared_exponential_kernel})
with $l=2.0$ as hyper-parameter.
In the left panel of Fig.~\ref{fig:gp_nn_comparison},
the drop of the cost function in
relation to the needed computational time is shown for the training
for each of the
\ac{NN} and \ac{GPR} machines.
Whereas training a \ac{NN}
converges after several thousand seconds, the \ac{GPR} approach
reaches the same level of accuracy within seconds.

Another difference in the implementation of the two methods lies
in the relative time each machine learning approach requires
to evaluate the function ${(x,
v_x)}^{\no{NHIM}}(\vec{y}, \vec{v_y}, t)$ for some input of the bath
coordinates.
For simplicity, we report the internal run time reported by the same
computer and call it the \emph{function call time}.
Precision for \ac{GPR} is achieved by adding more
training points and therefore decreasing the $95\,\%$ region of the posterior.
But adding more training points leads to an increasing dimension of the matrices
discussed in Sec.~\ref{kernel_methods}. Consequently,
the function call time required to evaluate Eq.~\eqref{gp_mu}
increases linearly with the number of
training points
as shown by the red curve in the right panel
of Fig.~\ref{fig:gp_nn_comparison}.
In contrast, increasing the number of training points does not increase the
evaluation time for the \ac{NN}, assuming its layer geometry does not
change.
Thus evaluating a small \ac{NN} can be significantly faster than
\ac{GPR}, although it might have taken a long time to train the net.

Based on these three observations,
we now have guidelines for choosing between
the \ac{NN} and \ac{GPR} machines for continuously interpolating the
time-dependent \ac{NHIM} or the related \ac{DS}:
(i) In cases when the function call time exceeds the additional training
time of the neural network compared to \ac{GPR}, it is computationally
more efficient to use the neural network approach.
A simple example would be
a system with time-periodic external driving (and without friction).
If the focus is on calculating rates for different initial ensembles of
particles under the influence of
the same external force, the \ac{NHIM} and
therefore the \ac{DS} is always the same. Consequently, it is more efficient to
use \acp{NN} so as to benefit from the short function call time.
(ii) On the other hand, in cases when the same ensemble
is used to calculate reaction rates
for a system
under influence of a different external driving or in cases with
non-periodic driving, the training time for each \ac{NN} can easily be
higher than the overall function call time to propagate the ensemble and
calculating reaction rates.
It is then more efficient to use \ac{GPR}.
Another example is the
future application of the methods presented here on thermal systems which
experience nontrivial friction and fluctuations.
In these systems, there will be a different \ac{NHIM}
(and \ac{DS}) for each sequence of random numbers modeling the fluctuations in a
Langevin-type approach, and each sequence will require separate training.
The training time can then easily exceed the function call time.
\EDITS{(iii) A third guideline can be formulated in terms of how many
  training points are needed for a given method. Using \acp{NN},
  millions of training points can be fit easily, whereas with GPR,
  the function call time is rather high even for just thousands of training points,
  and its use to fit millions of training points is hopeless.
  This comparison can also be seen exactly in the opposite way, as
  prior knowledge on the object to be interpolated---here the NHIM---can
  be included in the GPR method by choosing an appropriate kernel function.
  This usually reduces the amount of training data needed significantly and
  provides an advantage of GPR over the NN approach.
  If only a few training points are available (for example if training data
  is hard to obtain), GPR is still capable of interpolating
  these few points when provided a suitable kernel function,
  at least one with respect to optimization of the cost function.
}

\section{Application to a two-dimensional system}
\label{sec:application}

In this section, we assess the performance of the methods summarized above
for the central application of this work.
That is, the determination of the time-dependent NHIM and the associated rate
for the driven model potential \eqref{eq:potential}
for the same parameters as used in Sec.~\ref{multidim_sys}.
In the previous section, we focused on the accuracy as determined by the
cost function as one would typically use in training a \ac{NN}
or \ac{GPR} machine.
However, in our specific application,
we also have an external criterion at our disposal that allows us to assess the
accuracy of the \ac{NHIM}.
Namely, the \ac{DS} attached to the
\ac{NHIM} should by construction be recrossing-free.
For a numerical test of this criterion, a large ensemble of trajectories can be
launched in the neighborhood of the barrier and propagated forward in time
until it leaves this neighborhood. In the process, the number of crossings and
recrossings of the \ac{DS} can be monitored.
\EDITS{Beyond the decrease in the cost function,
  the demand of the DS to be as recrossing free as possible is
 another---and in our case more relevant---indicator of a successful
training process.}

To obtain a stringent test of the
\ac{DS}, it is advantageous to launch trajectories close to the stable
manifolds. These trajectories will remain in the neighborhood of the barrier for
a long time and therefore have plenty of opportunity to recross a \ac{DS} of low
quality.
A detailed description of such tests for the NHIM
constructed by other approaches, together with numerical
results, can be found in Refs.~\citenum{hern17h,hern18c} for various
test systems.
In what follows, we test the \ac{NN} and \ac{GPR} machines
relative to this criteria.

\subsection{Representation of the \ac{NHIM} by \acp{NN} or \ac{GPR}}

For both \acp{NN} and \ac{GPR}, training points are computed
in the range $y \in [-4, 4]$ and $v_y \in [-8, 8]$.
The \ac{NN} is
trained on an equidistant grid of $25$ points in $y$,  $40$ points in $v_y$
range and $40$ points in $t$, sampled over one period, i.e., in the interval
$[0, 2]$.
The structure of the net uses three input neurons for the variables
$y$, $v_y$, and $t$, four hidden layers with $40$, $40$, $40$, and $10$ neurons,
respectively, and one output neuron representing $x^{\mathrm{NHIM}}$.
While the
\ac{NN} discussed in Sec.~\ref{comparison} was written from scratch in our group, here
we resort to an efficient implementation using
TensorFlow.\cite{tensorflow2015-whitepaper}
To improve the periodicity of the resulting
function $x^{\mathrm{NHIM}}$, the grid of training points is tripled to cover a
full time span from $t\in [0, 6]$. All in all, we use $120\,000$ training
points taking advantage of the periodicity of the driven system, so just
$40\,000$ points have to be obtained using the binary contraction
method explained in Sec.~\ref{multidim_sys}. After training for $50\,000$
epochs, the training cost has decreased to $2.06\times 10 ^{-7}$ and the
verification cost to $2.15\times 10^{-7}$.
\EDITS{Although using so many training points might seem to be a bit conservative here,
  we did so as to ensure, that the likelihood of single, localized artifacts
  in the representation of the DS is reduced. Such artifacts,
  which are hidden by averaging over a large amount of accurately inferred data points
  in the loss function, will cause spurious recrossings to the trajectories.
This would lead to errors in the corresponding rates which goes against
our primary reason for calculating the non-recrossing \ac{DS}.
  Thus the large number of points---though not fully optimized---was chosen in
order to ensure not only the proper construction of the manifold but also to
satisfy our more stringent criterion for a reduced number of recrossings.}

For the \ac{GPR} representation of the \ac{NHIM}, we used $6000$
training points in total on equidistant grids chosen at $30$ time steps in the
range $t\in[-0.5, 2.4]$ and in the same $y$ and $v_y$ ranges as for the
\ac{NN}.
The time range is more than the full period of $T=2$ of the time-dependent
potential~\eqref{eq:potential} in order to improve periodicity.
Training was done using a squared exponential kernel like in
Eq.~\eqref{squared_exponential_kernel}, but with different characteristic length
scales $l$ in any direction $(y, v_y, t)$.
Training was carried out for various values of the hyperparameters on a
logarithmic grid, and the set that led to the smallest verification cost was
retained. The final values of the hyperparameters that best reflect the length
and time scales on which the \ac{NHIM} moves, were $\sigma_y\approx 0.705$,
$\sigma_{v_y} \approx 0.790$ and $\sigma_t \approx 0.505$.
\EDITS{Note, that the equidistant grids are chosen in order to provide no initial bias,
and thus focus on whether the different approaches are feasible. Choosing
non-equidistant grids may be a subject for future optimization of the training process.}

\begin{figure}[t]
  \centering
  \includegraphics[width=.90\figurewide]{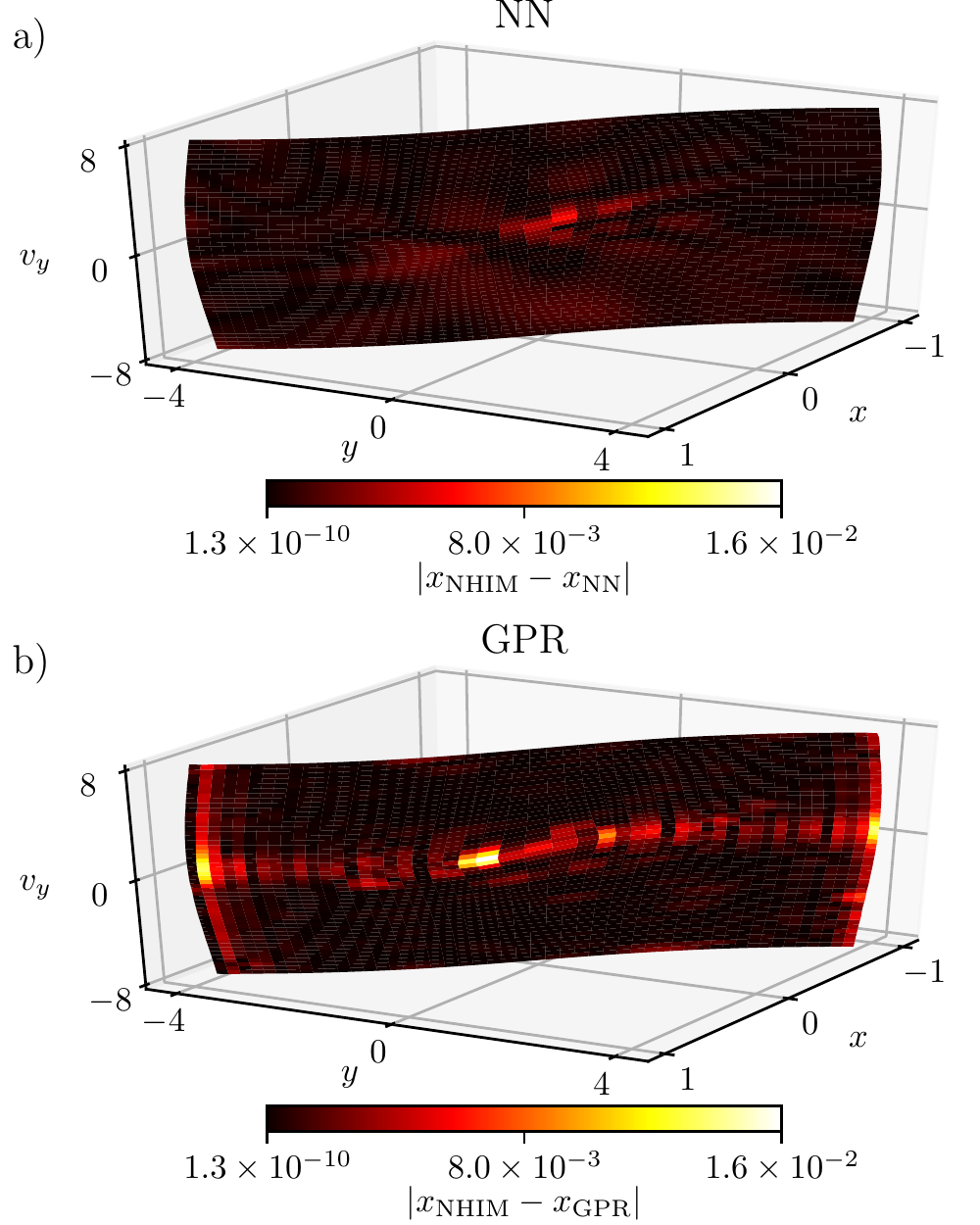}
\caption{%
  Estimation of the error of the \ac{NN} (a) and \ac{GPR}
  (b) machines in representing the \ac{NHIM}. Here, the values $x_{\mathrm{NHIM}}(y,
  v_y, 0)$ for a grid of $50\times 50$ points in the ranges $y \in [-4, 4]$ and
  $v_y \in [-8, 8]$ are obtained with an accuracy of $10^{-12}$ using the binary
  contraction method explained in Sec.~\ref{multidim_sys}. The color of the
  curved surface connecting these points represents the pointwise difference to
  the representation of the \ac{NHIM} by one of the two interpolation methods
  discussed here.}\label{fig:surface}
\end{figure}

To assess the accuracy of the representation obtained from either the
\ac{NN} or the \ac{GPR}, we computed the position of the NHIM
at a fixed time for a grid of $50\times 50$ additional points in the
ranges $y\in [-4, 4]$ and $v_y \in [-8, 8]$ with an accuracy of $10^{-12}$ using
the binary contraction method.
Afterwards, the values $x\sno{NN}(y,
v_y, 0)$ and $x\sno{GPR}(y, v_y, 0)$
were obtained for both trained machines
of the \ac{NHIM} at all of these $50\times 50$
verification points. Fig.~\ref{fig:surface} shows the location of the \ac{NHIM}
at time $t=0$ (in a three-dimensional projection of the four-dimensional
phase space). Very similar figures have been obtained for different
times. The color of the surfaces represents the error of the machine learning
representation of the \ac{NHIM} compared to the numerically exact result.

The maximum error of the \ac{NN} representation is $6.1\times 10^{-3}$ as
seen in Fig.~\ref{fig:surface}(a).
It is nearly a factor of three smaller than the
maximum error for \ac{GPR} in Fig.~\ref{fig:surface}(b).
Large errors of the \ac{NN}
occur just in a limited region near $y=v_y=0$, whereas the \ac{GPR}
has large errors in a band-like structure near $v_y=0$. The maximum
difference could be lowered by e.\,g.~using more training points.
But this will
come with the burden of longer training times for neural networks or longer
function call times for \ac{GPR}.
A different approach, to be pursued in the future,
might be the use of a non-equidistant grid of training points.
By increasing the density of points in the regions where the quality of the
representation is poor and decreasing it elsewhere, it should be possible to
lower the maximum error while keeping the total number of training points
constant.

As noted above, the \ac{GPR} interpolates the given training points
exactly, whereas the \ac{NN} attempts to minimize the training cost,
i.e., the mean square error, averaged over all training points, without bringing
it to zero. It now appears that the \ac{NN} produces a smoother
representation of the NHIM, with moderate errors everywhere, whereas the
\ac{GPR} yields very small errors near the training points at the expense of
larger errors elsewhere.

Note that a \ac{NN} typically has intrinsic hyper-parameters like the
activation function or the learning rate. Although adjusting these parameters is
not free, we neglect this adjustment time in our discussion, since in our
experience, these parameters have to be adjusted only once for the general
problem, but can be kept constant while just varying potential parameters as,
e.\,g.,~the amplitude of the moving saddle.

\begin{table}[t]
  \caption{Training and function call times
for the use of \ac{NN} and \ac{GPR} to learn and access
the same NHIM for the model problem discussed in the text.}
\label{tab:times}
  \begin{tabular}{lrr}
  \toprule
    & NN & GPR \\
  \midrule
  obtaining training data & $3636\,\no{s}$ & $545\,\no{s}$ \\
  training & $4220\,\no{s}$ & $9\,\no{s}$ \\
  optimize hyper-parameters & \textendash~~~~ & $2238\,\no{s}$ \\
  \midrule
  total training time & $7856\,\no{s}$ & $2792\,\no{s}$ \\
  \midrule
  \midrule
  function call time & $14\,\upmu\no{s}$ & $253\,\upmu\no{s}$ \\
  \bottomrule
  \end{tabular}
\end{table}

Finally, we comment on the numerical effort required by both methods, as
represented by the computing times listed in Table~\ref{tab:times}.
These times depend
critically on, e.\,g., the hardware used or the desired accuracy, and should be
understood as a rough guidance only. Computing all $40\,000$ training points for
the neural network took $3636\,\no{s}$, whereas obtaining $6\,000$ points for
the \ac{GPR} took only $545\,\no{s}$. (This corresponds to $0.091\,\no{s}$
on average per training point.) The process of training the neural net for
$50\,000$ epochs took $4220\,\no{s}$.
\ac{GPR}, by contrast, could
be trained in only $9\,\no{s}$. However, this training had to be repeated many
times in order to optimize the hyperparameters. In total, the process took
$2238\,\no{s}$. In summary, the neural network was was ready to use after
$7856\,\no{s}$ (about $131\,\no{min}$) the \ac{GPR} about
$2792\,\no{s}$ (about $47\,\no{min}$). Afterwards, the evaluation of a single
value of $x^{\textrm{NHIM}}$ took $14\,\upmu\no{s}$ for the neural network and
$253\,\upmu\no{s}$ for \ac{GPR}. Thus, although training in \ac{GPR}
is much faster, its function call time is by a factor of $18$ higher.
As we concluded at the end of Sec.~\ref{sec:machine},
depending on how many function calls are required after training, either method
can be more computationally efficient: The \ac{NN} becomes advantageous if
more than about 21 million evaluations are made.

\subsection{Relative accuracy of the rate constants:
\acp{NN} vs.~\ac{GPR}}

\begin{figure}[t]
  \centering
  \includegraphics[width=1.0\figurewide]{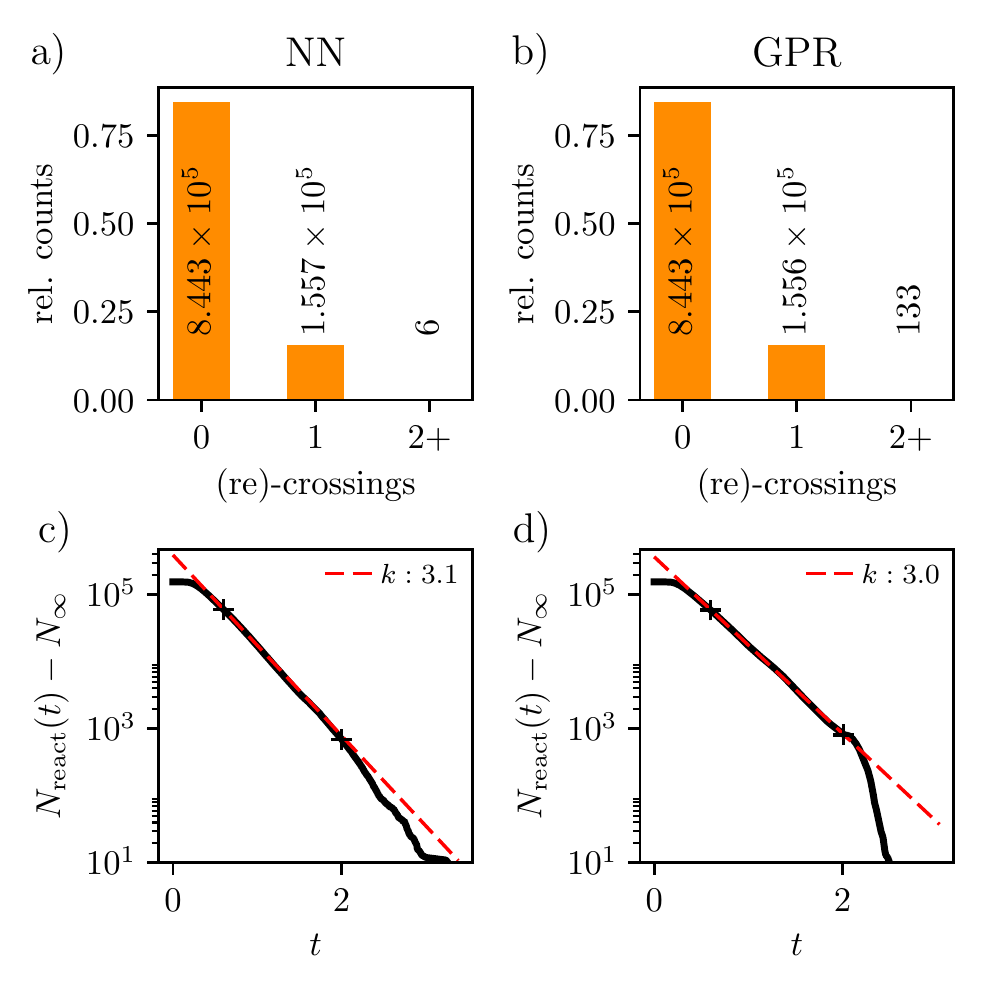}
\caption{%
Number of trajectories in the reactive ensemble crossing the
\ac{DS} 0, 1 or more times, as determined using
a \ac{NN} and \ac{GPR} in panels (a) and (b), respectively.
As the associated \ac{DS}
should be recrossing free, the number of trajectories with more than one
crossing (2+) gives an estimate of the  error in the computation. Whereas the
$y$-axis is in relative counts, the numbers on the bars are absolute numbers.
The corresponding
time-dependent traces of the reactant population $N\sno{react}(t)$
shifted by its infinite time limit are shown in
panels (c) and (d).
The rate constant $k$ \EDITS{given in units of a dimensionless inverse time}
can by found from the linear fit
(shown with a dashed red line) in the steady-state regime
according to Eq.~\eqref{rate_constant}.
}
\label{fig:hist_rate}
\end{figure}

To obtain a rate constant for a reaction over the time-dependent
barrier of potential~\eqref{eq:potential}, we propagate an ensemble of particles
initialized on the reactant side and monitor the reaction of particles to the
product side.
The ensemble is set up with $N\sno{react}(t=0) = 10^6$ particles
at $x_0 = -0.2$ with $y$ on the minimum energy path of
potential~\eqref{eq:potential} at time $t_0 = 0$.
The velocities $(v_x, v_y)$ of
these particles are sampled from a thermal distribution with $k\sno{B}\,T =
0.5$.

Propagating the full ensemble using a standard velocity-Verlet integrator, we
monitor possible crossings of the \ac{DS} and the related crossing times of each
particle. Ideally, each trajectory should either descend on the reaction side of
the barrier without crossing the \ac{DS} or cross it at exactly once and then
descend on the product side. In practice, because of inaccuracies in how the
\ac{DS} is represented, we expect a small number of trajectories to recross the
\ac{DS}.

A histogram with the number of crossings  is shown in Fig.~\ref{fig:hist_rate}
(a) for the \ac{NN} and in Fig.~\ref{fig:hist_rate}
(b) for the \ac{GPR} approach.
In both cases approx.~$16\,\%$ of all particles are
reactive and cross the \ac{DS} once from the reactant to the product side.
Nearly all other particles are non-reactive. The number of trajectories that
cross the \ac{DS} more than once is very small in both cases, though
the \ac{GPR} still produces a factor $22$ more recrossings than the
\ac{NN}.
This difference arises because the \ac{NN} represents the
\ac{NHIM} more accurately than \ac{GPR} as shown in
Fig.~\ref{fig:surface}.

Rates are obtained \EDITS{in dimensionless units} using the procedure described in
Sec.~\ref{subsec:rate_calculation}.
For the \ac{NN} approach in Fig.~\ref{fig:hist_rate} (c),
we obtain a rate constant of $k\sno{NN} = 3.1$
compared to the rate constant of $k\sno{GPR} = 3.0$ for the \ac{GPR}
approach in Fig.~\ref{fig:hist_rate} (d).
\EDITS{As discussed in Secs.~\ref{subsec:rate_calculation} and \ref{multidim_sys}, these rate
constants are given in units of a dimensionless inverse time. The time-scale
is given by a period with $T=2$ of the oscillation with frequency $\omega_x = \pi$
in Eq.~\eqref{eq:potential}.}
Although these constants are very similar,
$k\sno{NN}$ is presumed to be more precise
because the \ac{DS} of the \ac{NN} approach exhibits fewer recrossings.
Note that differences between the reactant populations computed by the two
methods arise once the populations have decayed to approximately the number of
recrossing trajectories in \ac{GPR}.

\section{Conclusion and outlook}
\label{conclusion}

In this paper, we have presented several techniques
to obtain the underlying geometric structures needed to obtain rate constants
for reactions over time-dependent driven barriers in the context of \ac{TST}.
Specifically, we focused on the determination of the
\ac{NHIM} which is
generally a multidimensional, time-dependent object located in phase space,
to which a recrossing-free \ac{DS} can be attached,
and with respect to which the TST rate is obtained.
As we restrict ourselves to rank-1 saddles,
the procedures presented here can in principle
be carried out for any
multidimensional system with one unstable degree of freedom and an
arbitrary number of stable bath degrees of freedom.
Here, one needs to keep in mind
that in higher-dimensional systems not only
does the propagation of trajectories
become more complex and computationally more challenging, but also the \ac{NHIM}
itself scales modestly with the dimensionality of the
equations of motion. Depending on how one chooses
to construct the \ac{DS}, the number of
points needed to characterize the underlying
\ac{NHIM} may scale exponentially with increasing dimensionality.

A central result of this work is the extension of recent approaches
to address reactions of increasing dimensionality.
We elaborate on the use of
our recently developed
\emph{binary contraction method}\cite{hern18g}
for the accurate non-perturbative determination of points on the NHIM.
In the case of time-periodically driven Hamiltonian systems,
this allowed us to extend the one-dimensional \ac{TS} trajectory
to multi dimensions by defining it to be the unique periodic
orbit that remains in the vicinity of the saddle.

A second central result of this work is the elaboration and extension of
our recent work on \acp{NN}.\cite{hern18c}
Here we show the applicability of the \ac{NN} to construct the
NHIM and rates for a two-dimensional system.
We also demonstrated that a different machine learning approach,
the \ac{GPR}, is also effective.
These two approaches have competing efficiencies in training and
use.
Together, they offer a powerful combination for obtaining
high-accuracy smooth \acp{NHIM} for barriers with rank-1
saddles in higher dimensions.
Indeed, Kamath et al\cite{carrington18}
recently compared both of these methods for the first time in the
context of constructing potential energy surfaces.\cite{carrington06}
Although their work has a different objective than the \acp{NHIM} obtained
here, we likewise
obtained similar conclusions regarding the advantages and
disadvantages of the two methods \EDITS{with respect to
the determination of the target surface}.
\EDITS{Carrington and coworkers\cite{carrington06}
employed a secondary criterion for evaluating the quality of their
computed surface. Namely,
the GPR outperforms NN when fitting potential energy surfaces to
obtain vibrational spectra.
In our case, we see in Fig.~\ref{fig:hist_rate} that the
total number of recrossings for the DS obtained by GPR is significantly
higher than that found by the NN. Following this additional
---and in our case more relevant---requirement to the
DS that it be recrossing-free, we conclude that for our applications the NNs are more
useful in the sense that they provide higher accuracy for an indirect
metric not used as part of the cost function while training.
We emphasize that we do not claim our setup to be fully optimized in
terms of net structure or the training process,
but it is the best we found using comparable approaches.
}

For the determination of the rate,
the computation of the required reaction
coordinates of the \ac{DS} from scratch (using e.\,g.~the binary
contraction method) whenever it is needed would be prohibitively
expensive, as the position of the \ac{DS} must be compared to the
instantaneous position of each propagated particle. Having a
continuous representation of the \ac{DS} as a function of bath modes and time
reduces this numerical effort drastically.
This is where the use of \ac{NN} and \ac{GPR} machine-learning techniques
represent a key advance by way of providing a suitable
\emph{smooth} surfaces.

It remains to show that these approaches are effective
for the determination of rate constants in systems with more
complex energy landscapes or external potentials.
This includes aperiodic external driving, potentials with multiple
saddles, dissipative terms and thermal noise.
In addition,
although these methods are developed for in principle arbitrary dimensional
problems, everything done so far was an application to
two- and three-dimensional model systems~\cite{hern17h,hern18c}. So one aspect
for a future work will be to test these methods on really high
dimensional systems with tens of degrees of freedom to answer the
question if for increasing dimensionality really an exponential increase
in the obtained number of points on the \ac{NHIM} is necessary or if
these points can be chosen more efficiently by improving the machine
learning procedures.

An application for future work will lead away from model systems
towards real systems like the LiCN $\leftrightarrow$ LiNC isomerization reaction,
to which an analytical approximation to the molecular
potential~\cite{essers1982scf} as well as to the molecular
dipole moment~\cite{brocks1984binitio}
is known.
Another possible application includes the ketene
isomerization\cite{gezelter1995,hern13c,wiggins14b}
which has received increased attention since the initial work
by some of us\cite{hern16d}
using the Lagrangian descriptor to obtain the \ac{NHIM} and the associated
rate constants.
The determination of the full-dimensional \ac{NHIM} in both of these cases remains
a challenge that may be addressable using the emerging approaches
presented here.

\begin{acknowledgement}

The German portion of this collaborative work was partially supported by
Deutsche For\-schungs\-ge\-mein\-schaft (DFG) through Grant No.\ MA1639/14-1. The US
portion was partially supported by the National Science Foundation (NSF) through
Grant No.~CHE 1700749. AJ acknowledges the Alexander von Humboldt Foundation,
Germany, for support through a Feodor Lynen Fellowship. MF is grateful for
support from the Landesgraduiertenf\"orderung of the Land Baden-W\"urttemberg.
This collaboration has also benefited from support by the European Union's
Horizon 2020 Research and Innovation Program under the Marie Sklodowska-Curie
Grant Agreement No.~734557. Surface plots
in Figs.~\ref{fig:crosses_with_surface} and \ref{fig:ts_traj}
have been made with the \emph{Mayavi}
software package~\cite{ramachandran2011mayavi}.

\end{acknowledgement}

\bibliography{p98long}

\end{document}